\documentclass[aps,pra,twocolumn]{revtex4-2}
\usepackage{graphicx} 
\usepackage{amsmath}
\usepackage{amssymb}
\usepackage{amsfonts}
\usepackage{amsthm}
\usepackage{braket}
\usepackage{changes}
\usepackage{bbm}
\usepackage{appendix}
\usepackage{kantlipsum}
\usepackage{subcaption}
\usepackage{adjustbox}
\usepackage{wrapfig}
\usepackage{hyperref}
\usepackage{bbm}
\usepackage{orcidlink}
\newtheorem{result}{Result}

\newcommand{\average}[1]{
\text{E}\left\{#1\right\} 
}

\newcommand{\Prob}[1]{
\text{Pr}\left\{#1\right\} 
}

\def\Cproduct{\C_{\text{p}}}

\newcommand{\EPE}{\text{EPE}}
\newcommand{\BayesRisk}{\text{L}^{\!*}}
\newcommand{\BayesRiskMin}{\BayesRisk_{\min}}
\newcommand{\BayesRiskMinLocal}{\BayesRisk_{\min,l}}
\def\Bayesriskaverage{\overline{\BayesRisk}_{\!\!\min}} 
\def\Bayesriskaveragelocal{\overline{\BayesRisk}_{\!\!\min,l}} 
\newcommand{\BayesClass}{g^*}
\newcommand{\ConditionalExp}{f^*}
\newcommand{\ConditionalExpOpt}{\ConditionalExp_{\boboptimalMeasurement}}
\newcommand{\QuadratEntropy}{\text{H}^{q}}
\newcommand{\Entropy}{\text{H}}

\def\quadraticentropyaverage{\overline{\QuadratEntropy}_{\!\!\min}}

\newcommand{\BinaryEntropy}[1]{\text{h}\!\left[#1\right]}
\newcommand{\certainty}{\mathcal{C}}
\newcommand{\certaintyOpt}{\mathcal{C}^*}

\newcommand{\QBER}{\epsilon}

\newcommand{\probError}[1]{
\text{L}\left(#1\right) 
}
\def\rateboundBB{K_{\text{BB84}}}
\def\centerSteeringEllipsoid{\vec{c}_{\text{se}}}
\def\inferenceVariance{\Delta}
\def\inferenceVarianceMin{\inferenceVariance_{\min}}
\def\inferenceVarianceMinLocal{\inferenceVariance_{\min,l}}
\def\inferenceVarianceMinAverage{\overline{\inferenceVariance}_{\min}}
\def\inferenceVarianceMinLocalAverage{\overline{\inferenceVariance}_{\min,l}}
\newcommand{\sharpObs}[1]{\mathcal{O}_{#1}}
\def\densityMatricesSpace{\mathcal{B}_1^+}

\def\boboptimalMeasurement{\vb^*}

\newcommand{\va}{\vec{a}}
\newcommand{\vb}{\vec{b}}
\newcommand{\C}{\textbf{C}}
\newcommand{\tA}{\vec{t}_A}
\newcommand{\tB}{\vec{t}_B}

\newcommand{\BlochSphere}{\text{B}_1}
\newcommand{\BlochBall}{\text{B}_{\leq 1}}

\newcommand{\Tr}[1]{
\text{Tr}\left[#1\right] 
}

\newcommand{\boptQuadraticEntropy}{
\vec{b}_{opt}
}

\def\devetakWinterK{K}
\def\ratebound{\tilde{\devetakWinterK}}
\def\rateboundOpt{\devetakWinterK^*}

\DeclareMathOperator*{\argmax}{arg\,max}
\DeclareMathOperator*{\argmin}{arg\,min}

\def\CLocalchannels{\C_{lc}}
\def\tAlc{\vec{t}_{A,\text{lc}}}
\def\tBlc{\vec{t}_{A,\text{lc}}}

\def\adcFunc{c_{\text{ad}}}
\def\pa{p_A}
\def\pb{p_B}
\def\mw{\mathbbm{w}}
\def\tAadc{\vec{t}_{A,\text{ad}}}
\def\tBadc{\vec{t}_{B,\text{ad}}}
\def\Cadc{\C_{\text{ad}}}

\def\errormeasuresAverage{\overline{\errormeasures}}
\def\errormeasures{U}

\usepackage{xspace}
\newcommand{\predictability}{unpredictability\xspace}
\newcommand{\Predictability}{Unpredictability\xspace}

\def\FHaar{F_{\text{Haar}}}

\def\FDos{F_{2}^{\text{CJWR}}}

\def\FTres{F_{3}^{\text{CJWR}}}

\begin{document}

\title{From top quarks to enhanced quantum key distribution: A Framework for Optimal Predictability of Quantum Observables}

\author{Dennis I. Mart\'inez-Moreno\,\orcidlink{0000-0001-9583-4873}}\affiliation{Departamento de Física Teórica, Atómica y Óptica, Universidad de Valladolid, 47011 Valladolid, Spain}
\author{Miguel Castillo-Celeita\,\orcidlink{0000-0003-2905-0389}}\affiliation{Departamento de Física Teórica, Atómica y Óptica, Universidad de Valladolid, 47011 Valladolid, Spain}
\author{Diego G. Bussandri,\,\orcidlink{0000-0002-1266-5750}}\email{diegogaston.bussandri@uva.es}\affiliation{Departamento de Física Teórica, Atómica y Óptica, Universidad de Valladolid, 47011 Valladolid, Spain}


\begin{abstract}
Predicting the outcomes of quantum measurements is a cornerstone of quantum information theory and a key resource for quantum technologies. Here, we introduce a comprehensive framework for quantifying the predictability of measurements on a bipartite quantum system using error measures inherited from statistical learning theory: the Bayes risk and inference variance. We derive analytical expressions for the optimal measurement that minimizes the prediction error for any arbitrary observable and any two-qubit state. We establish a direct, quantitative link between the ability to surpass the fundamental limit of local unpredictability and the presence of Einstein-Podolsky-Rosen steering. 
Additionally, by optimizing measurement choices according to the minimal Bayes risk, we propose a modified entanglement-based quantum key distribution protocol achieving higher secure key rates than the standard BB84 protocol, demonstrating enhanced resilience to noise. 
We apply our framework in two scenarios: perfect Bell states affected by local amplitude-damping noises, and top-antitop quark pairs produced in high-energy colliders. Our work offers a novel perspective on quantum correlations, connecting statistical inference, fundamental quantum phenomena, and cryptographic applications.

\end{abstract}

\maketitle

\section{Introduction}\label{sec:introduction}

The ability to predict the outcomes of quantum measurements lies at the heart of both foundational questions in quantum mechanics and practical applications in quantum information~\cite{coles2017}. In a bipartite scenario involving parties Alice and Bob sharing a quantum state $\rho_{AB}$, Bob can attempt to predict the outcome of Alice's measurement on her subsystem by performing a local measurement on his. The efficacy of this prediction is a direct reflection of the nature and strength of the correlations present in $\rho_{AB}$. 
In statistical learning theory, predictions are ruled by error measures such as the \textit{Bayes risk}, which quantifies the minimum probability of classification error, and the \textit{inference variance}, which measures the minimal expected prediction error~\cite{hastie_elements_nodate,cover_thomas2006}.

The study of Bayes risk in this quantum context represents, to the best of our knowledge, a novel contribution to the field. On the other hand, the inference variance was first introduced in Ref.~\cite{cavalcanti2009} under the name optimal inference variance in the context of uncertainty relations based on standard deviation.

In this work, we employ the Bayes risk and inference variance to construct a comprehensive framework for quantum predictability on the grounds of statistical learning theory. We first address two central questions: For a given measurement Alice performs, what is the optimal measurement Bob can choose to minimize his prediction error? What fundamental quantum resource enables Bob to make better predictions (on average) than any strategy restricted to local information on Alice's system alone? We answer these questions by deriving analytical solutions for the optimal measurements and the corresponding minimized prediction errors for arbitrary two-qubit states. Our key finding reveals a connection between statistical predictability and a fundamental quantum correlation: for Bell-diagonal states, the ability to exceed the \textit{local} predictability threshold is shown to be equivalent to the criterion for Einstein-Podolsky-Rosen steering. 

Next, through the Bayes risk optimization, we improve the performance of quantum key distribution protocols by proposing a modified entanglement-based protocol where Bob adaptively selects his measurement bases to minimize the Bayes risk for each of Alice's incompatible observables. We show that this approach yields higher asymptotic secure-key rates compared to the standard BB84 protocol~\cite{bennett2014}.

The averaged minimal measures and the modified entanglement-based protocol are particularly investigated when local amplitude-damping noises affect a maximally entangled Bell state~\cite{bussandri2024}, and an imperfect source such as top-antitop quark pair states produced in hadron colliders.

Top-antitop quark pairs are interesting emerging systems in quantum information~\cite{afik_quantum_2022,afik_quantum_2023,atlas2024,cheng2025,afik2025}. Since the pioneering work on this topic~\cite{afik_quantum_2022}, there has been an increased interest in exploring various quantum information quantities for these systems. Recent investigations have addressed phenomena such as entanglement and Bell non-locality~\cite{cheng2025}, reporting one of the highest-energy observations of entanglement in top-antitop quark pair production from proton-proton collisions at the Large Hadron Collider (LHC)~\cite{atlas2024}. Correlations such as quantum discord and steering have also been explored for this system~\cite{afik_quantum_2023,han2025}. The reader may consult Ref.~\cite{afik2025} for a roadmap of the future developments in this topic. 

Our work thus extends this line of reasoning by analyzing the capabilities of these quantum states in colliders as potential resources of entanglement-based quantum key distribution protocols. High-energy colliders provide a unique opportunity to test quantum information principles at the highest energy scales ever achieved, opening a new frontier for quantum technologies.

\textit{Structure of the article}. First, in Sec.~\ref{sec:Prelim}, we introduce and define the Bayes risk and the inference variance. In Sec.~\ref{sec:Predictability of quantum observables}, we addressed the predictability problem for an arbitrary quantum system $AB$ by identifying the optimal observable of $B$ that minimizes the prediction error of an observable of $A$. The \predictability of $A$-observables is defined in Sec.~\ref{sec_overcoming_local_pred} as the average over all observables on $A$, obtained for each of the previous statistical measures, allowing us to characterize the local \predictability. In Sec.~\ref{sec:Bayes risk in entanglement-based QKD protocols} a modified entanglement-based key distribution protocol is proposed, taking into account the optimized Bayes risk. Each section includes applications for two different resource states: perfect Bell pairs affected by local amplitude-damping noises (Sec.~\ref{sec:noisemodel}); and top-antitop quark ($t\bar{t}$) pairs produced in colliders (Sec.~\ref{sec_top_quarks_predictability}).

\section{Preliminaries} \label{sec:Prelim}

\begin{figure*}
    \centering
    \includegraphics[width=0.8\linewidth]{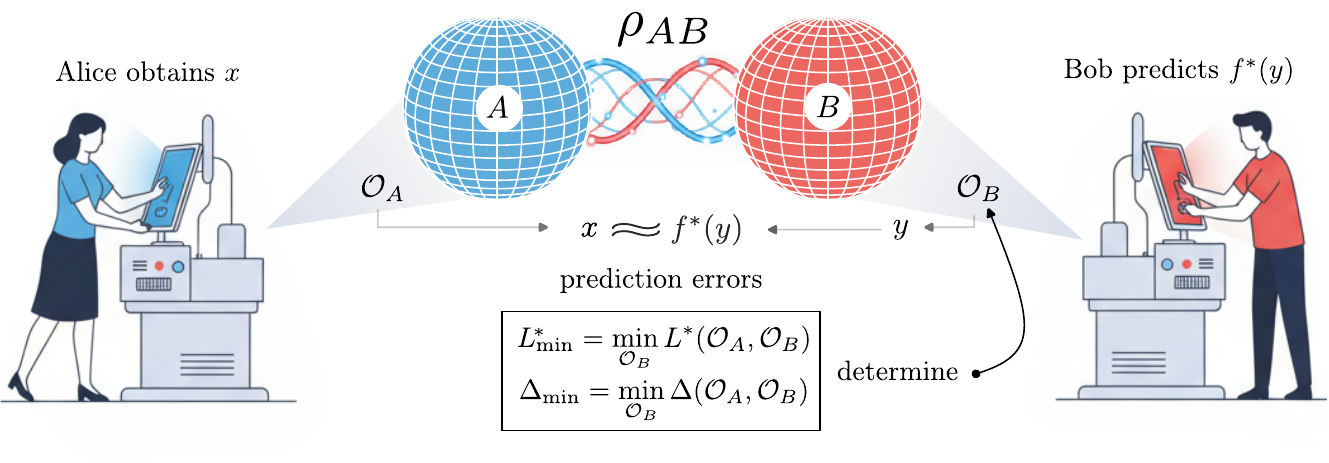}
    \caption{Diagram showing the predictability problem addressed in Sec.~\ref{sec:Predictability of quantum observables}, supporting Eqs.~\eqref{eq_bayes_risk_opt_problem} and~\eqref{eq_inference_variance_opt_problem}: Bob is intended to predict Alice's local measurement $x$ outcome with $\ConditionalExp(y)$. These local measurements, $\mathcal{O}_A$ and $\mathcal{O}_B$, are performed over a resource state $\rho_{AB}$. Bob's measurements are chosen to minimize the prediction error measures, the Bayes risk $\BayesRisk$~\eqref{eq:BayesRisk} and the inference variance $\inferenceVariance$~\eqref{eq:InferenceVariance}.}
    \label{fig_illustrative}
\end{figure*}


\subsection{Statistical learning basics: Bayes risk and conditional quadratic entropy}\label{sec:Statistical-Learning}

The ultimate goal of the statistical learning theory is to predict the value of a random variable $X$ may assume in an experiment, based on an observation of an auxiliary variable $Y$. If $X$ takes values in a discrete set $\mathcal{X}=\{0,\dots,|\mathcal{X}|-1\}$, there are two basic approaches: Classification and regression estimation. 

A \textit{classification} problem consists of predicting $X$ by guessing within the set $\mathcal{X}$, namely, the prediction is made through a \textit{classifier} $g:\mathcal{Y}\to\mathcal{X}$. In this case, an error occurs each time $g(y)\not= x$, with $(x,y)\in \mathcal{X}\times\mathcal{Y}=\mathcal{V}$, and thus the quality of the classifier can be assessed by:
\begin{align}
    \probError{g}=\Prob{g(Y)\not= X}=\average{\delta_{g(Y),X}},
\end{align}
where the average $\average{\cdot}$ is taken over the joint probability distribution of $X$ and $Y$, $P(X,Y)$, and $\delta_{x,y}$ is the Kronecker delta.

The Bayes classifier $\BayesClass$, defined as the optimal classifier $g$: $\BayesClass=\argmin_{g:\mathcal{Y}\to\mathcal{X}} \probError{g}$, is given by
\begin{align}
    \BayesClass(Y) = \argmax_{x\in\mathcal{X}} P(x|Y)= \argmax_{x\in\mathcal{X}}\frac{P(x,Y)}{P(Y)},
\end{align}
with $P(Y)=\sum_x P(x,Y)$.
Correspondingly, the \textit{Bayes risk} is the probability of error implied by $\BayesClass$, i.e. 
\begin{align}
    \BayesRisk=\probError{\BayesClass},
\end{align}
which is also known as Bayes probability of error or Bayes error~\cite{cover_thomas2006, devroye2013,hastie_elements_nodate}.

While the Bayes risk provides a useful way to evaluate the performance of a classifier, it does not capture the uncertainty inherent in the prediction process. In particular, it assumes that the random variable to be predicted takes determined classes, whereas there may be some degree of uncertainty or ambiguity in reality. A statistically relevant uncertainty measure in this context arises when we consider \textit{regression estimation}. 

The regression estimation problem involves predicting the random variable $X$ by some continuous function $f(y)$. To assess the regression optimality, the usual measure is the expected prediction error (EPE), or \textit{inference variance}~\cite{cavalcanti2009}
\begin{align}\label{eq:EPEregression}
    \EPE(f)=\average{[X-f(Y)]^2}.
\end{align}
The function $f$ optimizing this quantity is the \textit{conditional expectation}, namely, 
\begin{align}\label{eq:BayesProblemRegression}
    \ConditionalExp(Y)=\argmin \EPE(f)=\average{X|y}=\sum_{x\in\mathcal{X}} x \frac{P(x,Y)}{P(Y)},
\end{align}
leading to the \textit{optimal inference variance}:
\begin{align}\label{eq_inference_variance_def}
    \inferenceVariance = \EPE(f^*).
\end{align}
As we shall not consider any different function $f$ than $f^*$, we will refer to $\inferenceVariance$ as inference variance, skipping the `optimal' adjective.

For the two-class problem $\mathcal{X}=\{0,1\}$, or binary classification, the conditional expectation is closely related to the Bayes classifier:
\begin{align}
    \ConditionalExp(Y)&=\average{X|Y}=P(1,Y), \label{eq:CondExp} \\
    \label{eq:BayesClass}
    \BayesClass(Y)&=\left\{\begin{matrix}
        1 & \text{if } \ConditionalExp(Y)>1/2, \\
        0 & \text{otherwise}.
    \end{matrix}\right.
\end{align}
Namely, the solution to the Bayes problem, i.e. to find the optimal classifier $\BayesClass$, is ultimately defined by $\ConditionalExp$, which also optimizes the regression estimation problem, see Eq.~\eqref{eq:BayesProblemRegression}.

The inference variance can be expressed in terms of uncertainty by employing the \textit{quadratic entropy}~\cite{vajda1968}, 
\begin{align}
    \QuadratEntropy(p)=1-\sum_i p_i^2.\label{eq:QuadratEntropy_raw}
\end{align}
Specifically, $\inferenceVariance$ coincides with the conditional quadratic entropy~\cite{devijver1974}:
\begin{align}
    \inferenceVariance&=\sum_y P(y) \sum_x P(x|y)[1-P(x|y)]^2 \nonumber \\ 
    &=\sum_y P(y) [1-\sum_xP(x|y)^2],\label{eq:InferenceVariance}
\end{align}
providing an additional statistical significance to the inference variance. In addition, this quantity is also known as the \textit{logical entropy}~\cite{ellerman2013,ellerman2017,ellerman2018}, and the \textit{asymptotic nearest neighbor} error, i.e., the prediction error implied by the nearest neighbor rule for an infinitely large training set, among others~\cite{devroye2013}.

On the other hand, the Bayes risk admits the following forms:
\begin{align}
    \BayesRisk&=\average{[X-\BayesClass(Y)]^2}\nonumber\\
    &=\average{\min\{\ConditionalExp(Y),1-\ConditionalExp(Y)\}}\nonumber \\
    &= \frac{1}{2}-\frac{1}{2}\average{|2\ConditionalExp(Y)-1|}.\label{eq:BayesRisk}
\end{align}

Finally, a particular case to consider is defined by no auxiliary variable $\mathcal Y$. In this case, the prediction is fixed: One choice has to be taken among the possible values of $\mathcal X$. Again, the Bayes classifier $\BayesClass$ is given by the conditional expectation $\ConditionalExp$, which in this case reduces to the expectation \textit{per se}: $\ConditionalExp_{uc}=\average{X}=\sum_x xP(x)$. This case is equivalent to the general one when $\mathcal X$ and $\mathcal Y$ are uncorrelated variables, i.e., the joint probability distribution satisfies $P(X,Y)=P(X)P(Y)$. 

\section{\Predictability of quantum observables}\label{sec:Predictability of quantum observables}

In this Section, we will dive into the quantum observables' predictability through the classification and regression estimation problems defined in the previous Section.

Let us consider thus a bipartite quantum system $AB$, occupying the joint resource state $\rho_{AB}\in\mathcal{B}_1^+(\mathcal{H}_A\otimes\mathcal{H}_B)$, and let $\sharpObs{A}$ and $\sharpObs{B}$ two sharp observables of each system, respectively.

As outlined in Figure~\ref{fig_illustrative}, the predictability problem we shall deal with is to find the optimal observable $\sharpObs{B}$ minimizing the error in the prediction of $\sharpObs{A}$, following the statistical learning theory approach presented in Sec.~\ref {sec:Prelim}. By usual convention, we shall name as \textit{Alice} and \textit{Bob} the entities responsible for measuring $\sharpObs{A}$ and $\sharpObs{B}$, respectively. These measurements, represented respectively by projectors $\{M_x\}_x$ and $\{M_y\}_y$, define the two random variables $X$ and $Y$ denoting the local observable measurement outcomes, giving rise to the joint probability distribution $P(x,y)=\Tr{M_x \otimes M_y \ \rho_{AB}}$.

As we will work with finite-dimensional systems, two error measures can be defined, each related to the classification and regression estimation approaches:
\begin{align}
\BayesRiskMin(\sharpObs{A})&= \min_{\sharpObs{B}\in\densityMatricesSpace(\mathcal{H}_B)} \BayesRisk(\sharpObs{A},\sharpObs{B}), \label{eq_bayes_risk_opt_problem}\\
\inferenceVarianceMin(\sharpObs{A})&=\min_{\sharpObs{B}\in\densityMatricesSpace(\mathcal{H}_B)} \inferenceVariance(\sharpObs{A},\sharpObs{B}),\label{eq_inference_variance_opt_problem}
\end{align}
where we highlight the dependence of the error measures on the local observables. This section outlines the central problem we address.

In the case of two-qubit resource states $\rho_{AB}$, we were able to solve the optimization of both measures analytically, expressing the resulting quantities in terms of the elements defining the Fano form of $\rho_{AB}$,
\begin{align}
    [\C]_{ij}&=\Tr{(\sigma_i\otimes\sigma_j) \rho_{AB}}, \label{eq_Fanoform_C}\\
    [\tA]_i &= \Tr{(\sigma_i\otimes\mathbbm{1}) \rho_{AB}},\label{eq_Fanoform_tA}\\
    [\tB]_j &= \Tr{(\mathbbm{1}\otimes\sigma_j) \rho_{AB}}.\label{eq_Fanoform_tB}
\end{align}

Let us consider thus arbitrary local observables $\mathcal{O}(\vec{a})=\vec{a}\cdot\vec{\sigma}_A$ and $\mathcal{O}(\vec{b})=\vec{b}\cdot\vec{\sigma}_B$. Notice that the variables $X$ and $Y$ can take values in $\{0,1\}$, representing the \textit{up} and \textit{down} states along a particular direction, respectively. The joint probability distribution is given by,
\begin{align}\label{eq:JoinProb}
    P_{\va,\vb}(x,y)=\Tr{M_x(\vec{a})\otimes M_y(\vec{b}) \ \rho_{AB}},
\end{align}
where $\{M_z(\vec{x})\}_{z=0}^1$ are the corresponding eigenvectors of $\mathcal{O}(\vec{x})$.

The following two subsections are set to establish the optimized Bayes risk and the inference variance, i.e. solving Eqs.~\eqref{eq_bayes_risk_opt_problem} and~\eqref{eq_inference_variance_opt_problem}.

\subsection{Bayes Risk}\label{sec:BayesRisk}

The following result provides the minimal Bayes risk associated with the classification problem of predicting an arbitrary local observable in system $A$, by using measurements of local observables in $B$, for an arbitrary two-qubit quantum state $\rho_{AB}$.

\begin{result}\label{res:Bayesrisk}
The minimal Bayes risk $\BayesRisk$ of making a decision $\BayesClass$, see Eqs.~\eqref{eq:BayesRisk} and~\eqref{eq:BayesClass}, respectively, as prediction of the local measurement of the observable $\mathcal{O}(\va)=\va\cdot\vec{\sigma}$, given a measurement of $\mathcal{O}(\vb)$ over system $B$, for an arbitrary two-qubit quantum state $\rho_{AB}$, is
\begin{align} \label{eq:BayesRiskResult}
    \BayesRiskMin(\va)\!=\!\min_{\vb\in\BlochSphere}\BayesRisk(\va,\vb)\!=\!\begin{cases}
       \frac{1}{2}(1\!-\!|\va\cdot\tA|)\! &\! |\C^\intercal\va|<|\va\cdot\tA| \\ 
        \frac{1}{2}(1\!-\!|\C^\intercal\va|)\! &\! \text{otherwise},
        \end{cases}
\end{align}
where $\C$ is the correlation matrix of $\rho_{AB}$, Eq.~\eqref{eq_Fanoform_C}, and $\tA$ the Bloch vector of the reduced state corresponding to system $A$, Eq.~\eqref{eq_Fanoform_tA}.
The optimal observable is $\mathcal{O}(\vb^*)=\boboptimalMeasurement\cdot\vec{\sigma}$ with, 
\begin{align}
\boboptimalMeasurement=\argmin_{\vb\in\BlochSphere}\BayesRisk=\C^\intercal\va/|\C^\intercal\va|.
\end{align} 
\end{result}
Appendix~\ref{app:BayesRiskOpt} provides the corresponding proof.

The previous result indicates that Bob's prediction $\BayesClass(y)$, Eq.~\eqref{eq:BayesClass}, is determined by the conditional expectation Eq.~\eqref{eq:CondExp}, for the optimal observable which can be written as:
\begin{align}\label{eq:CondExp2}
    \ConditionalExpOpt(y)=\frac{1}{2}\left\{1-\vec{a}\cdot\left[\frac{\vec{t}_A+(-1)^y (\textbf{C}\boboptimalMeasurement)}{2P_{\boboptimalMeasurement}(y)}\right]\right\},
\end{align}
where $P_{\boboptimalMeasurement}(y)=\frac{1}{2}[1+(-1)^y\boboptimalMeasurement\cdot\vec{t}_B]$ stands for the probability of obtaining $Y=y$, i.e. an outcome $y$ in the measurement of $\mathcal{O}(\boboptimalMeasurement)$, and $\tB$ the Bloch vector of the reduced density matrix of system $B$, see Eq.~\eqref{eq_Fanoform_tB}.

It is worth mentioning that if $|\C^\intercal\va|<|\va\cdot\tA|$, $\BayesRiskMin$ is independent of the information provided by measuring system $B$: $\BayesRiskMin$ does not depend on $\vb$ and Bob's decision $\BayesClass(y)$ is completely specified by the sign of $\va\cdot\tA$ [i.e. $\BayesClass(0)=\BayesClass(1)$]. Hence, the absolute values of $\C^\intercal\va$ and $\va\cdot\tA$ tell whether to decide using local information on $A$, or to prefer the correlations between the subsystems quantified by the correlation matrix $\C$. 

In the case in which the correlations are stronger, $|\C^\intercal\va|>|\va\cdot\tA|$, the Bayes risk reduces to the optimal quantum bit error rate (QBER) $\QBER$: The QBER is the probability that Alice and Bob's measurement outcomes disagree, indicating a mismatch between their outcomes~\cite{yin2017}. For arbitrary observable directions $\va$ and $\vb$, the QBER reads
\begin{align}\label{eq:QBER}
    \QBER(\va,\vb)=P_{\va,\vb}(0,1)+P_{\va,\vb}(1,0)=\frac{1}{2}(1-\vec{a}\cdot\textbf{C}\vec{b}),
\end{align}
therefore:
\begin{align}
    \QBER(\va,\vb)\geq \QBER(\va,\boboptimalMeasurement)=\frac{1}{2}(1-|\C^\intercal\va|),
\end{align}
which coincides with the optimized Bayes risk under the assumption $|\C^\intercal\va|>|\va\cdot\tA|$.

If the quantum state $\rho_{AB}$ does not involve any kind of correlations between $A$ and $B$, the resource state is $\rho_{AB}^p=\rho_{A}\otimes\rho_{B}$ with correlation matrix $\Cproduct = \tA\tB^\intercal$, and the joint probability distribution Eq.~\eqref{eq:JoinProb} results in a product of independent events; besides, 
\begin{align}\label{eq_product_states}
    |\Cproduct\va| = t_B |\va\cdot\tA|\leq |\va\cdot\tA|.
\end{align} 
Thus, the Bayes risk reduces to 
\begin{align}\label{eq_bayesriskmin_uncorrelated}
    \BayesRiskMinLocal(\va) = \frac{1}{2}(1-|\va\cdot\tA|), 
\end{align}
for all $\va\in \BlochSphere$. 

It is worth noting that the previous uncorrelated scenario, as mentioned in Sec.~\ref{sec:Prelim}, is equivalent to the non-auxiliary variable case which, in this context, it is usually referred to as \textit{local \predictability} scenario on $A$~\cite{costa2016,cavalcanti2009,jevtic2015} because there is no subsystem $B$.

\subsection{Inference Variance}\label{sec_conditional_quadratic_entropy_results}

\begin{figure*}
     \centering
     \begin{subfigure}[!]{0.4\textwidth}
         \centering
         \includegraphics[width=\textwidth]{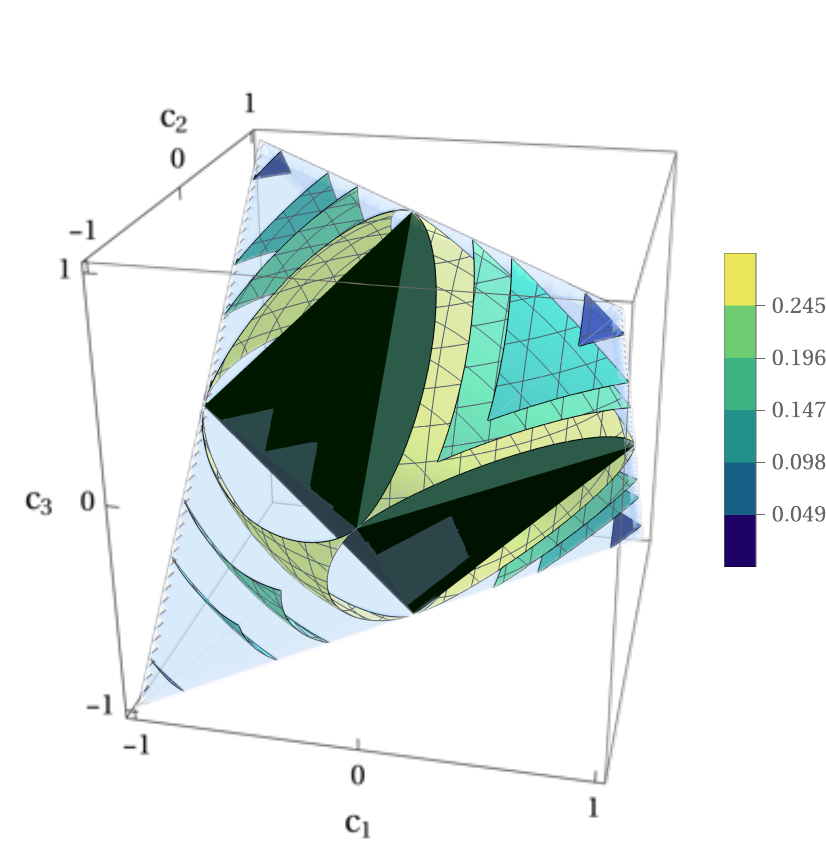}
         \caption{Bayes Risk, $\Bayesriskaverage$.}
         \label{fig_BD_predictability_bayes}
     \end{subfigure}
     \begin{subfigure}[!]{0.4\textwidth}
         \centering
         \includegraphics[width=\textwidth]{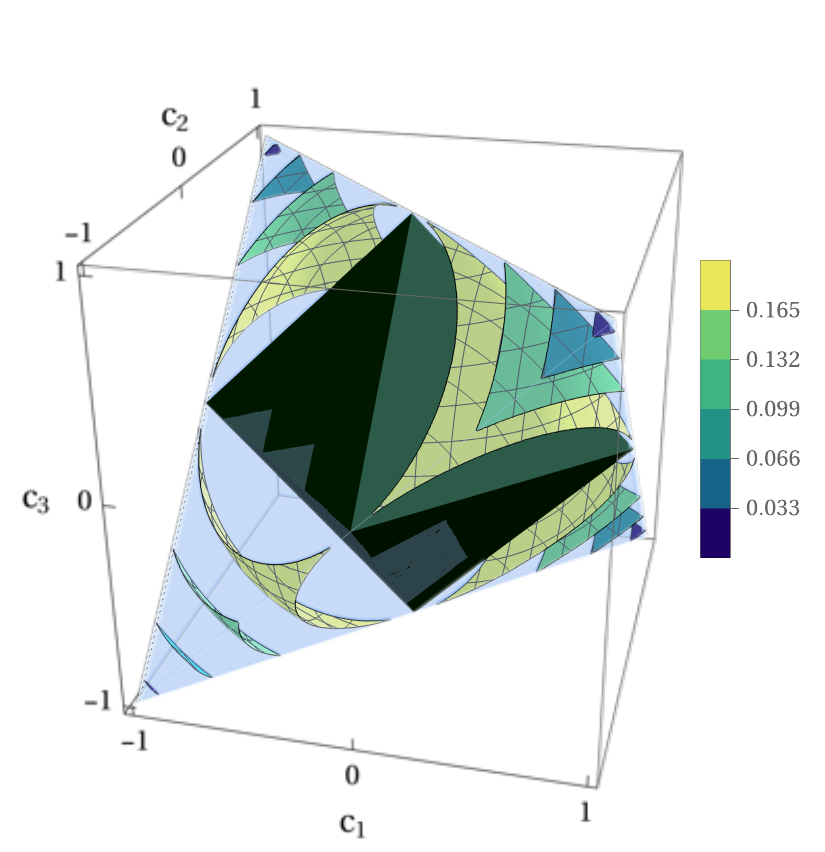}
         \caption{Inference variance, $\inferenceVarianceMinAverage$.}
         \label{fig_BD_inf_var}
     \end{subfigure}
            \caption{Contour plot of Bayes risk $\Bayesriskaverage$ [Eq.~\eqref{eq:BayesRiskAverage}] and inference variance $\inferenceVarianceMinAverage$ [Eq.~\eqref{eq:QEntropyAverage}], in the tetrahedron spanned by Bell-diagonal states. Separable states correspond to the octahedron, while the yellow contours denote the corresponding local \predictability thresholds: $\Bayesriskaverage=1/4$ and $\inferenceVarianceMinAverage=1/6$.}
        \label{fig_BD_predictability}
\end{figure*}

Let us consider now that Bob aims to approximate Alice's measurement result with a regression function. As mentioned in Sec.~\ref{sec:Statistical-Learning}, the usual error measure is $\inferenceVariance$, see Eq.~\eqref{eq:EPEregression}, where the function $\ConditionalExp$ stands for the conditional expectation, Eq.~\eqref{eq:BayesProblemRegression}. The joint probability distribution of the measurement outcomes in this case is given by Eq.~\eqref{eq:JoinProb}. The following result provides the analytical solution of Eq.~\eqref{eq_inference_variance_opt_problem}, i.e. the minimal inference variance and the optimal observable to be measured in system $B$.


\begin{result}\label{res:QuadratEntropy}
    The optimal inference variance $\inferenceVariance$, Eq.~\eqref{eq_inference_variance_opt_problem}, as given by the joint probability distribution Eq.~\eqref{eq:JoinProb}, results:
    \begin{align}
        \inferenceVarianceMin &= \min_{\vb\in\BlochSphere}\inferenceVariance = \frac{1}{4}\left(1-\certaintyOpt\right),\label{eq_quadraticEnt_opt} \\
        \certaintyOpt &= (1-|\tB|^2)|\va\cdot\centerSteeringEllipsoid|^2+|\C^\intercal\va|^2, \label{eq_predic_opt} 
    \end{align}
    where
    \begin{align}\label{eq_center_ellipsoid}
        \centerSteeringEllipsoid = \frac{\tA-\C\tB}{1-t_B^2}.
    \end{align}
    The optimal measurement observable is given by:
    \begin{align}
        \boptQuadraticEntropy=\frac{\C^\intercal\va-(\va\cdot\centerSteeringEllipsoid)\tB}{|\C^\intercal\va-(\va\cdot\centerSteeringEllipsoid)\tB|}. \label{eq_b_opt} 
    \end{align} 
\end{result}

The underlying calculations for the previous result are presented in Appendix~\ref{app:proof R1}.

The vector $\centerSteeringEllipsoid$, appearing naturally when optimizing the inference variance, corresponds to the centroid of the \textit{quantum steering ellipsoid}~\cite{jevtic2015}, which geometrically represents the set of conditional states that Bob can steer on Alice's subsystem by measuring sharp observables (interior points can be reached with generalized measurements). We refer the reader to Appendix~\ref{app_steering} for a short compendium on steering ellipsoids and their associated inequalities.

It is important to mention that the $\inferenceVarianceMin$ is well-defined for all physical quantum states $\rho_{AB}$ (even for $|\tB|=1$) because of the positivity conditions~\cite{GAMEL2016}.


As mentioned in the previous Sec.~\ref{sec:BayesRisk}, the local \predictability implied by the inference variance can be obtained from Eq.~\eqref{eq_quadraticEnt_opt} by setting an uncorrelated state through $\Cproduct = \tA \tB^\intercal$. Thus:
    \begin{align}
        \inferenceVarianceMinLocal = \frac{1}{2}\left(1-|\va \cdot \tA|^{2}\right). \label{eq_predic_optlocal} 
    \end{align}


\section{Overcoming the local \predictability}\label{sec_overcoming_local_pred}

Let $\errormeasures$ be the optimal Bayes risk, Eq.~\eqref{eq:BayesRiskResult}, or inference variance, Eq.~\eqref{eq_quadraticEnt_opt}. Its average $\errormeasuresAverage$ over all possible sharp observables on $A$~\cite{holevo2019} defines a Haar measure integration, which in the Bloch sphere parametrization translates into the following surface integral:
\begin{equation}\label{eq_averaged_quantities}
\errormeasuresAverage(\tA, \tB, \C) = \frac{1}{4\pi} \int_{\BlochSphere} \errormeasures(\va,\tA, \tB, \C) \, \mathrm{d}S,
\end{equation}
where $dS$ represents the differential solid angle over the Bloch sphere $\BlochSphere$. We shall refer to $\errormeasuresAverage$ as the \textit{\predictability} of $A$-observables.

In this section, we aim to explore what kind of quantum correlations in the resource state are needed to improve the optimal local \predictability as measured by the previous two quantities, $\BayesRiskMin$ and $\inferenceVarianceMin$, see Eqs.~\eqref{eq:BayesRiskResult} and \eqref{eq_quadraticEnt_opt}, respectively. The local \predictability arises by setting an uncorrelated state, leading to $\errormeasuresAverage_l(\tA)=\errormeasuresAverage(\tA,\tB,\tA\tB^\intercal)$, see Sec.~\ref{sec:Prelim}. We aim thus to characterize the resource states for which $\errormeasuresAverage$ is less than the local \predictability threshold:
\begin{align}\label{eq_predictability_threshold}
    \min_{\tA\in\BlochBall} \errormeasuresAverage_l(\tA) > \errormeasuresAverage(\tA,\tB,\C).
\end{align}

\subsection{Average Minimal Bayes Risk}

The dependence of the Bayes risk $\BayesRiskMin$ on the absolute values of $\C^\intercal\va$ and $\va \cdot \tA$ makes it difficult to obtain analytical expressions for the average in Eq.~\eqref{eq_averaged_quantities} for general quantum states $\rho_{AB}$. However, we can solve the integral for quantum states satisfying
\begin{align}\label{eq_assumption}
    |\C^\intercal\va|>|\va \cdot \tA|, \ \text{ for all } \va\in\BlochSphere,
\end{align}
or the opposite case. The calculation of the local predictability threshold constitutes an example of the previous case: By setting $\Cproduct=\tA\tB^\intercal$, it follows Eq.~\eqref{eq_product_states}. Therefore, the Bayes risk is continuous in $\va$, and we can perform the integration analytically:
\begin{equation}\label{eq:BayesRiskAverageL}
\Bayesriskaveragelocal = \frac{1}{2}\left(1-\frac{|\tA|}{2}\right).
\end{equation}
See Appendix~\ref{app:AveragePredictability} for details about this calculation. The local \predictability threshold in this case reads 
\begin{align}\label{eq_minimalBayesRiskAverageL}
    \min_{\tA\in\BlochBall} \Bayesriskaveragelocal = \frac{1}{4}.
\end{align}

On the other hand, if we consider the case in Eq.~\eqref{eq_assumption}, meaning that always the measurement outcome in $B$ is taken into account to predict Alice's measurement result, see Sec.~\ref{sec:BayesRisk}, the \predictability can also be obtained analytically. Quantum states having maximally mixed reduced states (i.e. $\tA=\tB=\vec{0}$) serve as a particular example set satisfying the previous Eq.~\eqref{eq_assumption}. 

Within this assumption, we showed in Appendix~\ref{app:AveragePredictability} that the average of $\BayesRiskMin$ is determined by the Carlson symmetric elliptic integral $R_{G}$:
\begin{equation}\label{eq:BayesRiskAverage}
\Bayesriskaverage = \frac{1}{2}\left[1-|c_{1}|R_{G}\left(\frac{|c_{2}|^{2}}{|c_{1}|^{2}},\frac{|c_{3}|^{2}}{|c_{1}|^{2}},1\right)\right],
\end{equation}
being $\{c_{1}, c_{2}, c_{3}\}$ the coefficients $\C_d$ of the diagonal matrix corresponding to the singular value decomposition of $\C$. 

By mixing Eqs.~\eqref{eq_minimalBayesRiskAverageL} and \eqref{eq:BayesRiskAverage}, we conclude that quantum states overcoming the optimized local \predictability as quantified by the Bayes risk, satisfy:
\begin{align}\label{eq_condition_bayesrisk}
    \Bayesriskaverage\leq \frac{1}{4} \ \iff \frac{1}{2} \leq |c_{1}|R_{G}\left(\frac{|c_{2}|^{2}}{|c_{1}|^{2}},\frac{|c_{3}|^{2}}{|c_{1}|^{2}},1\right).
\end{align}

Interestingly, the previous condition is very well-known in the context of Einstein–Podolsky–Rosen (EPR)-steering: Eq.~\eqref{eq_condition_bayesrisk} coincides with the criterion for all sharp observables (see Appendix~\ref{app_steering}, Eq.~\eqref{eq_steering_ineq_haar_meas})
\begin{align*}
    2\pi<\int d\Omega\sqrt{\vec n \cdot\C \C^\intercal \vec n},
\end{align*}
that EPR-steerable states satisfy for Bell-diagonal states~\cite{jevtic2015}. This showcases the connection between steering and overcoming the local \predictability threshold.

Figure~\ref{fig_BD_predictability_bayes} displays the contour lines of $\Bayesriskaverage$ in the tetrahedron defining all physical Bell-diagonal states. The yellow surface indicates those states reaching the local \predictability threshold $\Bayesriskaverage=1/4$.

\begin{figure*}
    \centering
    \includegraphics[width=0.8\linewidth]{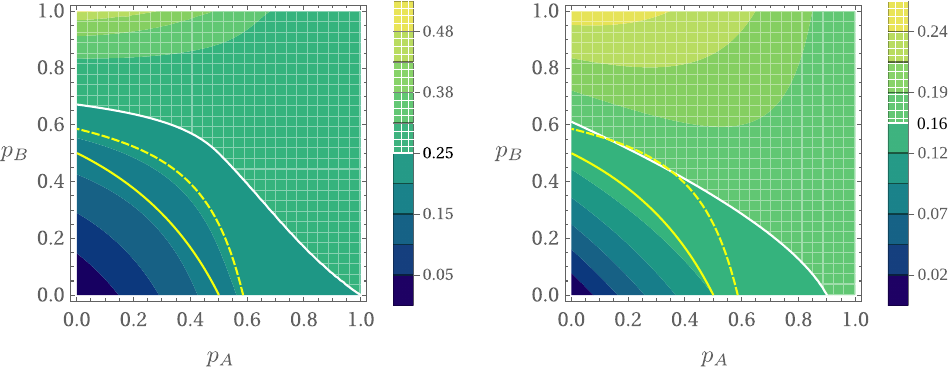}
    \caption{Contour lines of the average Bayes risk $\Bayesriskaverage$ (left), Eq.~\eqref{eq:BayesRiskAverage}, and inference variance $\inferenceVarianceMinAverage$ (right), Eq.~\eqref{eq:QEntropyAverage}, as functions of the local damping parameters $\pa$ and $\pb$. Their corresponding local predictability thresholds are shown as a continuous white lines and the steering inequalities $F_2^{\text{CJWR}}\leq1$ and $F_3^{\text{CJWR}}\leq1$ are also displayed as solid and dashed yellow lines, respectively.
    }
    \label{fig_amplitude_damp_unpredic}
\end{figure*}
\subsection{Average Minimal Inference Variance}

The \predictability, as measured by the average of the minimal inference variance $\inferenceVarianceMin
$, Eq.~\eqref{eq_quadraticEnt_opt}, can be calculated for arbitrary resource states:
\begin{align}\label{eq:QEntropyAverage}
   \inferenceVarianceMinAverage&= \frac{1}{4}\left[1-\frac{(1-t_B^2)|\centerSteeringEllipsoid|^2+\|\C\|^2}{3}\right],
 \end{align}
 with $\|\C \|^{2}= \text{Tr}[\C\C^{\dagger}]$ being the Hilbert-Schmidt inner product. We include the corresponding calculation details in Appendix~\ref{app:AveragePredictability}.
 
On the other hand, the average minimal inference variance when the subsystems are uncorrelated, $\Cproduct=\tA\tB^\intercal$, given in Eq.~\eqref{eq_predic_optlocal}, is determined by $\inferenceVarianceMinLocalAverage = \frac{1}{4}\left(1-\frac{t_A^{2}}{3}\right)$; thus, the local \predictability threshold corresponding to the inference variance is:
\begin{align}\label{eq:QEntropyAverageL}
\min_{\tA\in\BlochBall}\inferenceVarianceMinLocalAverage=\frac{1}{6}.
\end{align}

By having the analytical expression of the \predictability as measured by the inference variance, $\inferenceVarianceMinAverage$, we demonstrate (see Appendix~\ref{app_improving_LU_quadratic_entr}) that classical-quantum two-qubit states do not improve the local \predictability threshold, i.e. $\inferenceVarianceMinAverage \geq 1/6$. On the other hand, Bell-diagonal states leading to $\inferenceVarianceMinAverage < 1/6$ are equivalent to those satisfying the steering criterion for three observables $\FTres \leq 1$, see Eq.~\eqref{eq_steering_ineq_three_meas}, Appendix~\ref{app_steering}.

In Figure~\ref{fig_BD_inf_var}, we show the contour lines of $\inferenceVarianceMinAverage$ in the tetrahedron of Bell-diagonal states. The yellow surface represents states with $\inferenceVarianceMinAverage = 1/6$. Additionally, we observe that the Bayes risk provides a less restrictive condition for improving local predictability ($\Bayesriskaverage \leq 1/4$) compared to the inference variance. Specifically, any state satisfying $\inferenceVarianceMinAverage\leq 1/6$ will automatically satisfy the corresponding Bayes risk inequality.

\subsection{Noise model: Local amplitude-damping channels}
\label{sec:noisemodel}

Now, we will evaluate the previous averaged measures [Eq.~\eqref{eq_averaged_quantities}] for a resource state resulting from the action of local noises, represented by amplitude-damping channels, over a maximally entangled Bell state.

The action of two quantum channels $\mathcal{E}$ and $\mathcal{F}$ with affine decomposition $(A_{\mathcal{E}},\vec{b}_{\mathcal{E}})$ and $(A_{\mathcal{F}},\vec{b}_{\mathcal{F}})$~\cite{karol2017}, respectively, over a maximally entangled Bell state $\left|\Phi_k\right>\!\left<\Phi_k\right|$, results in the resource state $\mathcal{E} \otimes \mathcal{F}\left(\left|\Phi_k\right>\!\left<\Phi_k\right|\right)$. The corresponding Fano form [see Eq.~\eqref{eq_Fanoform}] is~\cite{bussandri2024}
\begin{align}\label{eq:noiseschannels}
\tAlc &= \vec{b}_{\mathcal{E},} \ \text{ } \ \tBlc = \vec{b}_{\mathcal{F}}, \\ 
    \CLocalchannels &= \left(\vec{b}_{\mathcal{E}}\vec{b}_{\mathcal{F}}^\intercal+ A_{\mathcal{E}} \mathbbm{w}_k A^\intercal_{\mathcal{F}}\right),
\end{align}
where $\mathbbm{w}_k$ stands for the correlation matrix of $\left|\Phi_k\right>\!\left<\Phi_k\right|$ in the Fano form~\eqref{eq_Fanoform}.

In the case of two local amplitude-damping channels $\mathcal{E}_{\text{ad}}$ and $\mathcal{F}_{\text{ad}}$, the resulting resource state 
\begin{align}\label{eq_resource_two_adc}
\rho_{AB}^{\text{ad}}=\mathcal{E}_{\text{ad}}\otimes\mathcal{F}_{\text{ad}}(\left|\Phi_k\right>\!\left<\Phi_k\right|),
\end{align}
is given by~\cite{bussandri2024}:
\begin{align}
\tAadc&=\pa\hat{k}, \ \tBadc=\pb\hat{k}, \label{eq:TWOADCReducedStates} \\
\Cadc&=\mw_1\text{diag}\{\adcFunc,\adcFunc,\adcFunc^2\}+\pa\pb\hat{k}\hat{k}^\intercal,\nonumber\\
\adcFunc&=\sqrt{\!(1\!-\!\pa)\!(1\!-\!\pb)},
\label{eq:TWOADCCOrrMat}
\end{align}
where $\pa$ and $\pb$ are the corresponding damping parameters in systems $A$ and $B$, respectively.

This noise model was extensively studied in the context of quantum teleportation, giving rise to a phenomenon called \textit{fighting noise with noise}~\cite{fortes2015}, and also experimentally investigated in Ref.~\cite{knoll2014}. Additionally, it constitutes an interesting state set to explore because, if $\pa \not= 1$ and $\pb\not= 1$, the resulting resource state is entangled for all damping parameters.


\subsubsection{\Predictability of $A$-observables for $\rho_{AB}^{\text{ad}}$}

Fig.~\ref{fig_amplitude_damp_unpredic} displays the contour lines of $\Bayesriskaverage$ and $\inferenceVarianceMinAverage$ as functions of the local damping parameters $\pa$ and $\pb$. The continuous white line indicates the corresponding local \predictability threshold. As shown, both measures exhibit asymmetric behavior where increasing noise in system $A$ is less detrimental than increasing noise in system $B$. Moreover, for this noise model, the inference variance imposes stricter requirements than the Bayes risk: if a state improves local predictability for the Bayes risk, it will also improve local predictability for the inference variance.

Regarding quantum correlations, as previously mentioned, $\rho_{AB}^{\text{ad}}$ is entangled for all $\pa\neq1$ and $\pb\neq1$. Moreover, the yellow lines indicate the steering inequalities $\FDos\leq 1$ (solid line, Eq.~\eqref{eq_steering_ineq_two_meas}) and $\FTres\leq 1$ (dashed line, Eq.~\eqref{eq_steering_ineq_three_meas}). Therefore, Fig.~\ref{fig_amplitude_damp_unpredic} demonstrates that in both cases, a substantial portion of entangled states fail to exceed the local \predictability threshold.

For the Bayes risk, all steerable states fall below the local predictability threshold. However, this does not hold for the inference variance, where we observe a region that exceeds its corresponding threshold.

\begin{figure*}
      \centering
      \begin{subfigure}[!]{0.325\textwidth}
         \centering
          \includegraphics[width=\textwidth]{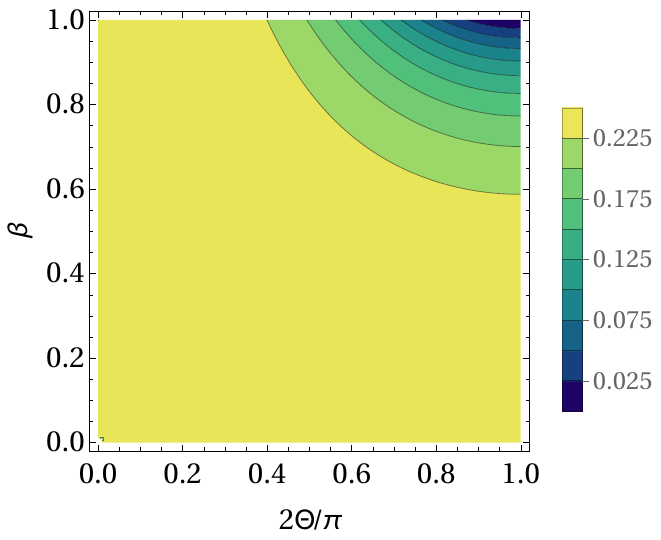}
          \caption{$w_{gg}=0$}
          \label{}
      \end{subfigure}
      \hfill
      \begin{subfigure}[!]{0.325\textwidth}
          \centering
          \includegraphics[width=\textwidth]{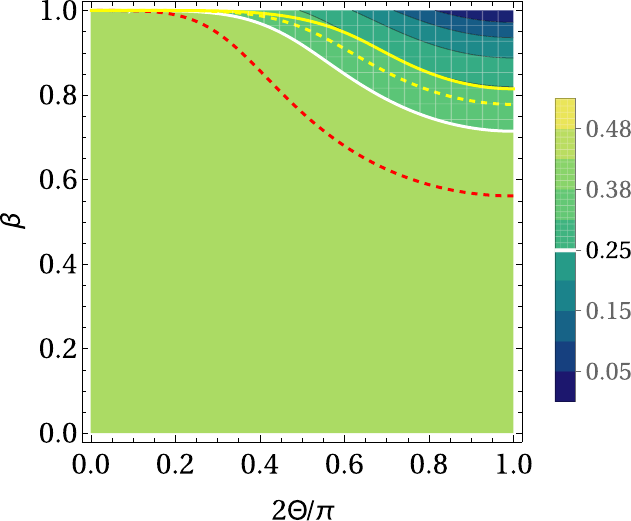}
          \caption{$w_{gg}=0.25$}
          \label{}
      \end{subfigure}
       \hfill
      \begin{subfigure}[!]{0.325\textwidth}
          \centering
          \includegraphics[width=\textwidth]{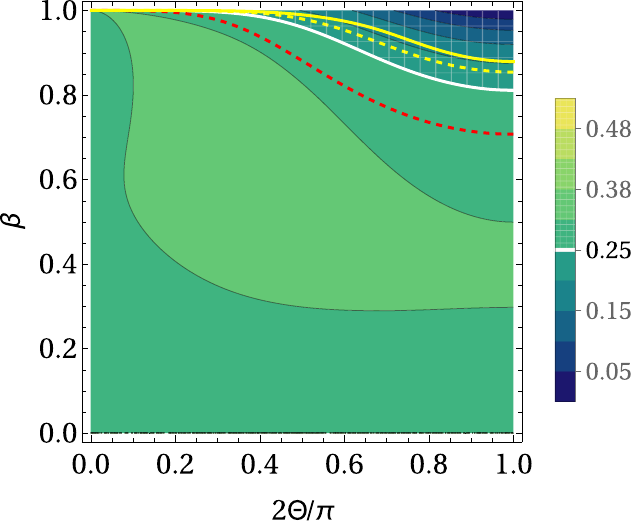}
          \caption{$w_{gg}=0.50$}
          \label{}
      \end{subfigure}
      \begin{subfigure}[!]{0.325\textwidth}
          \centering
          \includegraphics[width=\textwidth]{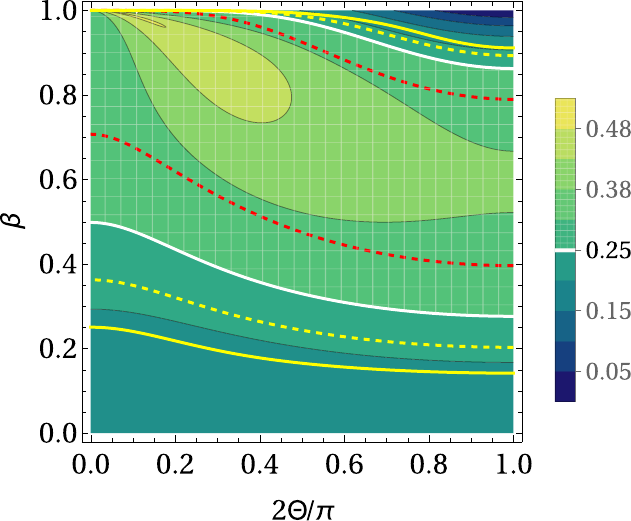}
          \caption{$w_{gg}=0.75$}
          \label{}
      \end{subfigure}
      \begin{subfigure}[!]{0.325\textwidth}
          \centering
          \includegraphics[width=\textwidth]{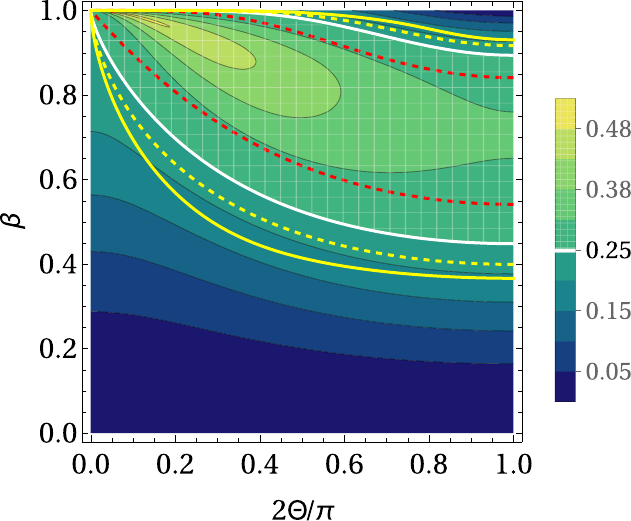}
          \caption{$w_{gg}=1$}
          \label{}
      \end{subfigure}
             \caption{Contour plot of the average Bayes risk $\BayesRiskMin$ for the statistical mixture of the $gg$ and $q\bar{q}$ processes, weighted by $w_{gg} \in [0,1]$, in the production of $t\bar{t}$ pairs. The local predictability threshold, $\Bayesriskaveragelocal = 1/4$, is shown as a continuous white line. Also shown are the three steering inequalities described in Appendix~\ref{app_steering}: $F_2^{\text{CJWR}} \leq 1$ (solid yellow line), $F_3^{\text{CJWR}} \leq 1$ (dashed yellow line), and $\FHaar \leq 1$ which coincides with the local predictability threshold (white line). The region where the resource state of the $t\bar{t}$-pair exhibits entanglement is bounded by dashed red lines.} 
           \label{fig:Bayesriskquarks}
 \end{figure*}

 \begin{figure*}
      \centering
      \begin{subfigure}[!]{0.325\textwidth}
          \centering
          \includegraphics[width=\textwidth]{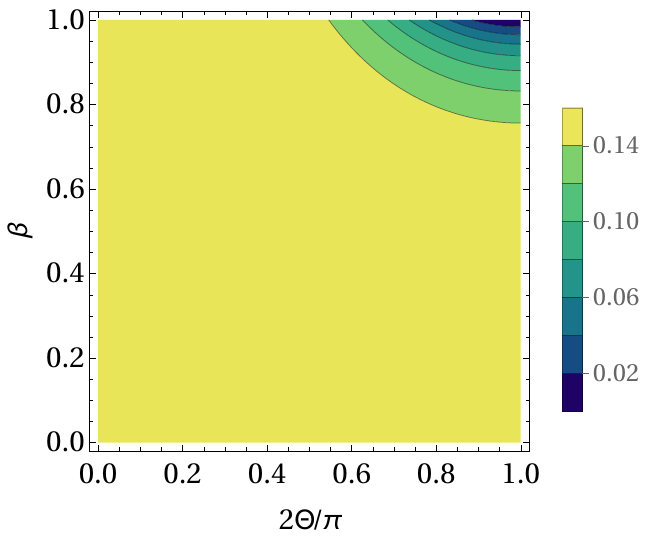}
          \caption{$w_{gg}=0$}
         \label{}
      \end{subfigure}
      \hfill
      \begin{subfigure}[!]{0.325\textwidth}
          \centering
          \includegraphics[width=\textwidth]{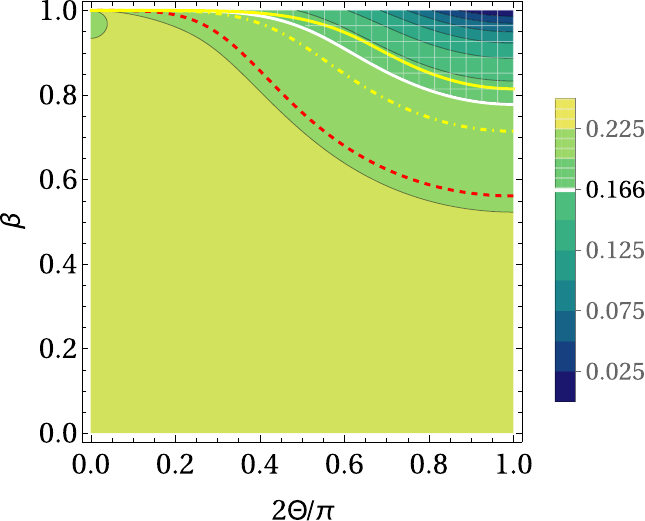}
          \caption{$w_{gg}=0.25$}
          \label{}
      \end{subfigure}
       \hfill
      \begin{subfigure}[!]{0.325
     \textwidth}
          \centering
          \includegraphics[width=\textwidth]{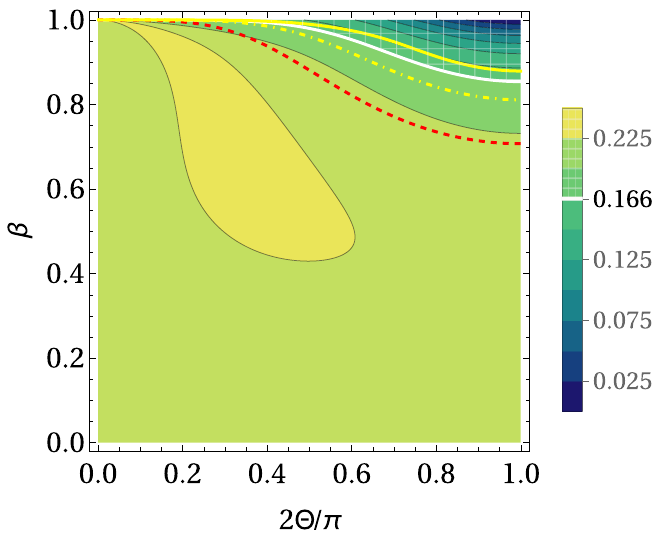}
          \caption{$w_{gg}=0.50$}
          \label{}
      \end{subfigure}
      \begin{subfigure}[!]{0.325
     \textwidth}
          \centering
          \includegraphics[width=\textwidth]{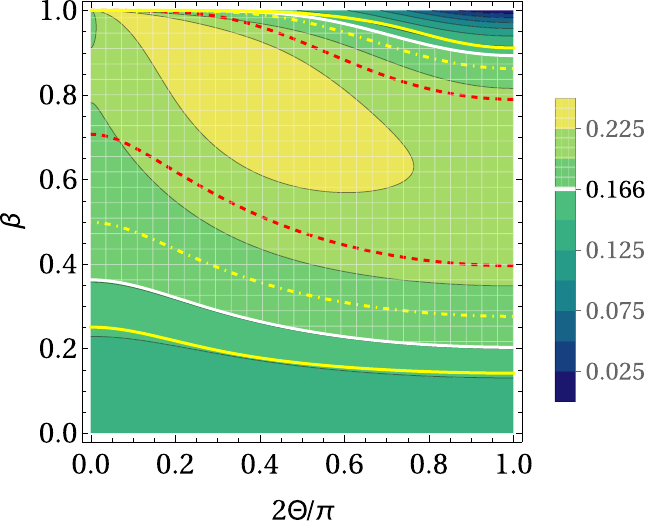}
          \caption{$w_{gg}=0.75$}
          \label{}
      \end{subfigure}
      \begin{subfigure}[!]{0.325
     \textwidth}
          \centering
          \includegraphics[width=\textwidth]{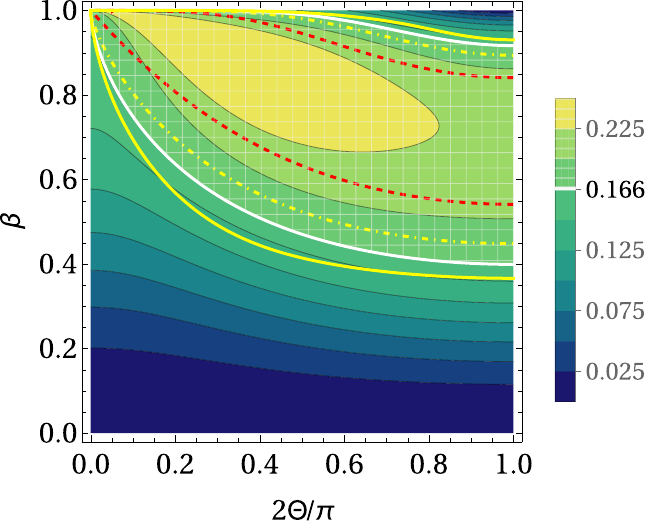}
          \caption{$w_{gg}=1$}
          \label{}
      \end{subfigure}
         \caption{
         Average inference variance, $\inferenceVarianceMinAverage$ for the statistical mixture of the $gg$ and $q\bar{q}$ processes, weighted by $w_{gg} \in [0,1]$, in the production of $t\bar{t}$ pairs. The local predictability threshold, $ \inferenceVarianceMinLocal = 1/6$, is shown as a continuous white line. Also shown are the three steering inequalities described in Appendix~\ref{app_steering}: $F_2^{\text{CJWR}} \leq 1$ (solid yellow line), $F_3^{\text{CJWR}} \leq 1$ which coincides with the local predictability threshold (white line), and $\FHaar \leq 1$ (yellow dash-dotted line). The region where the resource state of the $t\bar{t}$-pair exhibits entanglement is bounded by dashed red lines.
         }
         \label{fig:Entropyquarks}
 \end{figure*}

\subsection{Imperfect source: \Predictability in top quarks in QCD} \label{sec_top_quarks_predictability}

In this Section, we will focus on analyzing the \predictability for top quarks' quantum states in colliders, a recent topic mainly introduced in Ref.~\cite{afik_quantum_2022}.

The spin information of top quark and top antiquark pairs, also known as \textit{top-antitop} ($t\bar{t}$) pairs, can be encoded into a two-qubit state (see Appendix \ref{app:TopQuarks} for a short introduction to the topic). The production of these pairs occurs in proton-proton ($pp$) or proton-antiproton ($p\bar{p}$) collisions at a specific energy $\sqrt{s}$. At leading order perturbation theory, these pairs can be produced by two processes: Light quark-antiquark interaction ($q\bar{q}$), and gluon-gluon interaction ($gg$). Thus, the density matrix of the pair is a statistical mixture of the contributions from the previous processes, $q\bar{q}$ and $gg$, weighed by a parameter $w_{gg}$ (i.e. the probability of obtaining a top-antitop from $gg$ processes). Additionally, this state is diagonalized by the Bell basis when the center of mass frame is used to describe the dynamics. 

In summary, the spin quantum state of $t\bar{t}$-pairs results in,
\begin{align}\label{eq_top_quark_state}
    \rho^{t\bar{t}}_{AB}=\rho^{t\bar{t}}_{AB}(\beta, \Theta, \sqrt{s}),
\end{align}
which depends on $\beta$, the velocity of the pair, and $\Theta$, the production angle with respect to the beam direction. The reduced states are maximally mixed ($\tA=\tB=\vec{0}$), and the correlation matrix $\C_{t\bar{t}}$ is diagonal.

For simplicity, and because it provides a suitable approximation~\cite{afik_quantum_2022}, we assume that the weight probability is constant: $w_{gg}(\beta, \Theta, \sqrt{s})=w_{gg}$. Thus, the correlation matrix of $\rho^{t\bar{t}}_{AB}$ is given by
\begin{align}\label{eq:CorrelationMixture}
    \C_{t\bar{t}}(\beta, \Theta,w_{gg})=w_{gg}\C^{gg}(\beta, \Theta)+(1-w_{gg})\C^{q\bar{q}}(\beta,\Theta).
\end{align}
The matrix elements of $\C^{I}(\beta, \Theta)$ are specified in Appendix~\ref{app:TopQuarks}, Eqs.~\eqref{eq_qq_process} and~\eqref{eq_gg_process}.

\subsubsection{\Predictability of $A$-observables for $\rho_{AB}^{t\bar t}$}

We present the contour lines of the average Bayes risk ($\Bayesriskaverage$) and average inference variance ($\inferenceVarianceMinAverage$) in Figs.~\ref{fig:Bayesriskquarks} and \ref{fig:Entropyquarks}, respectively, with the corresponding local predictability thresholds shown as continuous white lines. We also delineate the phase space region $(\beta, \Theta)$ where the $t\bar{t}$-pair resource state exhibits entanglement (bounded by dashed red lines), together with the steering inequalities detailed in Appendix~\ref{app_steering}: $F_2^{\text{CJWR}}\leq1$ (yellow continuous line), $F_3^{\text{CJWR}}\leq1$ (yellow dashed line), and $\FHaar\leq1$ (yellow dash-dotted line).

We find that for pure quark-antiquark processes ($w_{gg}=0$), both measures fall below their corresponding local predictability thresholds, revealing an entire class of states that improve predictability. However, as the fraction of gluon-gluon processes increases, a large phase space region rapidly emerges where the state fails to improve local predictability, similar to the entanglement behavior presented in Ref.~\cite{afik_quantum_2022}.

We identify similar behavior patterns for both measures as we incorporate gluon-gluon processes: Near threshold production ($\beta=0$), states do not improve local predictability for $w_{gg}=0.25$ or $w_{gg}=0.5$; however, when $w_{gg}\geq 0.75$, a region emerges where they do improve predictability. Conversely, near $\beta\to1$ and $\Theta \to \pi/2$, predictability deteriorates with increasing $w_{gg}$, as the beneficial effects become confined to the vicinity of the point $(\beta,\Theta)\to (1,\pi/2)$.


\begin{figure*}
    \centering
    \includegraphics[width=.8\linewidth]{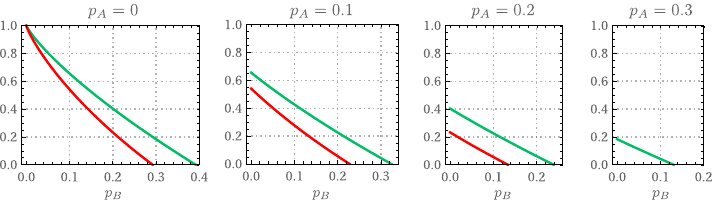}
    \caption{Devetak-Winter rates for BB84 protocol $\rateboundBB$, Eq.~\eqref{eq_devetak_winter_bb84}, (red line) and optimized modified protocol $\rateboundOpt$, Eq.~\eqref{eq_rate_bound_opt}, (green line) as functions of damping parameter $\pb$ for fixed values $\pa\in\{0,0.1,0.2,0.3\}$.}
    \label{fig_two_adc_qkd}
\end{figure*}

\begin{figure}
    \centering
    \includegraphics[width=.8\linewidth]{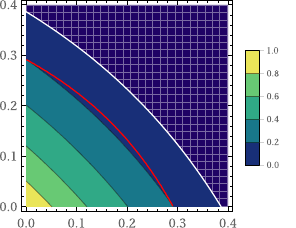}
    \caption{Contour lines of the optimized rate $\rateboundOpt$, Eq.~\eqref{eq_rate_bound_opt}, as a function of damping parameters $\pa$ and $\pb$. The red line indicates the BB84 security threshold ($\rateboundBB=0$), while the white line shows the modified protocol threshold ($\rateboundOpt=0$).}
    \label{fig_two_adc_qkd_cont}
\end{figure}

\section{Bayes risk in entanglement-based QKD protocols}\label{sec:Bayes risk in entanglement-based QKD protocols}

The \textit{BB84} protocol stands for the most typical example of a Quantum Key Distribution (QKD) protocol~\cite{ramona2021}. Its entanglement-based version involves Alice and Bob measuring the observables defined by the Pauli operators $\sigma_x$ or $\sigma_z$~\cite{yin2017}. When they measure the same observable, they retain the resulting bit; otherwise, they discard it. The collection of non-discarded bits constitutes the \textit{sifted key}.  See Appendix~\ref{app_qkd} for a summary on key distribution and the main security rate, known as the Devetak-Winter (DW) rate or asymptotic secure key rate.

The analytical expression of the DW rate for the BB84 protocol is
\begin{align}\label{eq_devetak_winter_bb84}
    \rateboundBB = 1-\BinaryEntropy{\QBER_z}-\BinaryEntropy{\QBER_x}
\end{align}
where $\QBER_z=\QBER(\vec{k},\vec{k})$ and $\QBER_x=\QBER(\vec{i},\vec{i})$ are the quantum bit error rates [see Eq.~\eqref{eq:QBER}] corresponding to observables $\sigma_z$ or $\sigma_x$, respectively, and $$\BinaryEntropy{p}=-p\log_2p-(1-p)\log_2(1-p),$$ is the binary entropy.

Ideally, in the absence of noise and eavesdropping, the error rate in the sifted key would be zero because, among other reasons, the optimal observable for Bob's measurement to maximize the coincidence with Alice's measurement result is identical to the observable Alice measures. In other words, the quantum transmission phase of the entanglement-based BB84 protocol is optimal for Bell states, in the sense that the observables to be measured by Alice and Bob ($\sigma_x$ and $\sigma_z$) lead to the minimal (in this case, null) error rate if there is no eavesdropping or imperfections in the generation or distribution of the resource state. However, in practice, errors may arise in the bit shared strings from environmental noise or because the source emitting states $\rho_{AB}$ is intrinsically imperfect.

Let us propose hence an entanglement-based \textit{modified} protocol based on performing measurement pairs, over $A$ and $B$, optimizing the Bayes risk (see Sec.~\ref{sec:BayesRisk}). We will see that, in principle, the secure key rate of this modified version improves that of the BB84 protocol, $\rateboundBB$. The fundamental assumption for this is that Alice and Bob rely on a characterization of the pre-distributed resource state, meaning they have reliable information about $\rho_{AB}$.

As mentioned before, the main protocol modification is taking the following two pairs of measurements: 1) $\mathcal{O}(\va_1)=\va_1\cdot\vec{\sigma}$ and $M[\boboptimalMeasurement(\va_1)]=\boboptimalMeasurement(\va_1)\cdot\vec\sigma$, and 2) $\mathcal{O}(\va_2)=\va_2\cdot\vec\sigma$ and $M[\boboptimalMeasurement(\va_2)]=\boboptimalMeasurement(\va_2)\cdot\vec\sigma$, respectively. Note that now Bob's measurements optimize the Bayes risks for each Alice observable; see Result~\ref{res:Bayesrisk}. Besides, let us consider that Alice's observables are incompatible: $\va_1\cdot\va_2=0$. Without loss of generality, from now on, we shall assume that the key is generated by just one of Alice's measurements, for example, $\mathcal{O}(\va_1)$. Finally, we fixed $\vec{a}_1$ and $\vec{a}_2$ by optimizing the following security rate bound based on each $\BayesRisk$.

\begin{result}\label{res:rate_Bound}
    The asymptotic secure key rate $\devetakWinterK(\va_1,\va_2)$ for the entanglement-based modified protocol defined above for Alice's incompatible measurements, $\mathcal{O}(\vec{a}_1)$ and $\mathcal{O}(\vec{a}_2)$, is lower bounded by:
\begin{small}
    \begin{align}\label{eq:devetak_winter_bb84_mod}
        \devetakWinterK(\va_1,\va_2) \geq 1 -  \BinaryEntropy{\BayesRiskMin\!(\va_1)}-\BinaryEntropy{\BayesRiskMin(\va_2)}=\rateboundOpt(\va_1,\va_2),
    \end{align}
\end{small}

\noindent being $\BayesRiskMin(\va_i)$ the optimized Bayes risk implied by measurements $\mathcal{O}(\va_i)$ and $M[\boboptimalMeasurement(\va_i)]$. Additionally,
if
\begin{align}\label{eq_rate_bound_opt}
    \max_{\va_1 \perp \va_2} \rateboundOpt(\va_1,\va_2)= \rateboundOpt(\va_1^*,\va_2^*),
\end{align}
it follows:
\begin{align}\label{eq:devetak_winter_bb84_mod_vs_bb84}
    \devetakWinterK(\va_1^*,\va_2^*)\geq \rateboundBB.
\end{align}
\end{result}
See appendix \ref{app:Rate_bound_demonstration} for the proof.

\subsubsection{Noise model: Local amplitude-damping noises, $\rho^{\text{ad}}_{AB}$}

In this Section, we analyze the entanglement-based QKD modified protocol proposed in Sec.~\ref{sec:Bayes risk in entanglement-based QKD protocols}, when the resource state is affected by two local amplitude-damping noises, i.e. the resource state of the protocol is $\rho^{\text{ad}}_{AB}$, c.f. Eq.~\eqref{eq_resource_two_adc}, and determined by local damping parameters. 

Fig.~\ref{fig_two_adc_qkd} shows the DW rate corresponding to the BB84 protocol $\rateboundBB$ (red line), and our rate bound $\rateboundOpt(\va_1^*,\va_2^*)$ for the modified version, optimized over all incompatible measurements (green line). Additionally, we calculate numerically the actual DW rate for the protocol defined by measurements $\va_1^*$ and $\va_2^*$,  $\devetakWinterK(\va_1^*,\va_2^*)$, by employing the conic optimization procedure introduced in Ref.~\cite{andres2025}. This figure exhibits plots for fixed values of the damping parameter in $A$, $\pa\in\{0,0.1,0.2,0.3\}$. Firstly, we see that the bound $\rateboundOpt(\va_1^*,\va_2^*)$ coincides with $\devetakWinterK(\va_1^*,\va_2^*)$ (green line). Besides, our modified protocol improves the security rate considerably: The BB84 protocol turns out to be insecure around $\pa\approx0.3$ or $\pb\approx0.3$, while our modified version does around $\pa\approx 0.4$ or $\pb\approx 0.4$. This fact is showcased in Fig.~\ref{fig_two_adc_qkd_cont}, where we plot the contour lines of $\rateboundOpt(\va_1^*,\va_2^*)$ as a function of the damping parameters. The red line stands for the region where $\rateboundBB=0$, while the white line indicates $\rateboundOpt(\va_1^*,\va_2^*)=0$.


\begin{figure*}
      \centering
      \begin{subfigure}[!]{0.3\textwidth}
          \centering
          \includegraphics[width=\textwidth]{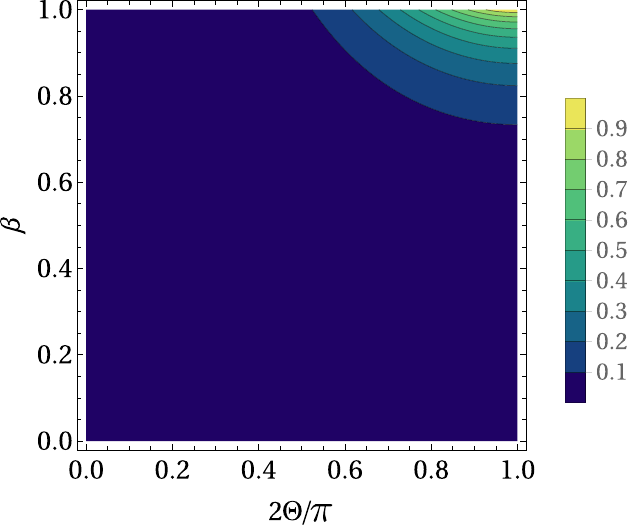}
          \caption{$w_{gg}=0$}
         \label{}
      \end{subfigure}
      \hspace{1cm}
      \begin{subfigure}[!]{0.3\textwidth}
          \centering
          \includegraphics[width=\textwidth]{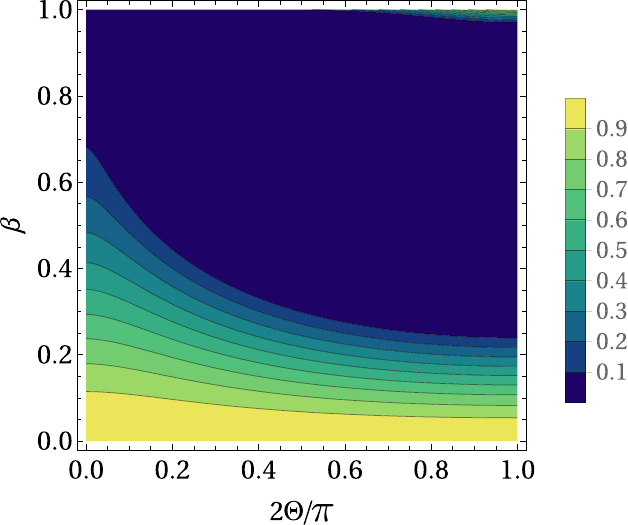}
          \caption{$w_{gg}=1$}
          \label{}
      \end{subfigure}
         \caption{Devetak-Winter security rate $\rateboundBB$ for entanglement-based BB84 protocol using $\rho_{AB}^{t\bar t}$ resource states. Left panel shows pure quark-antiquark processes ($w_{gg}=0$), right panel shows pure gluon-gluon processes ($w_{gg}=1$).} 
         \label{fig_sec_rate_bb84}
 \end{figure*}

 \begin{figure*}
      \centering
      \begin{subfigure}[!]{0.3\textwidth}
          \centering
          \includegraphics[width=\textwidth]{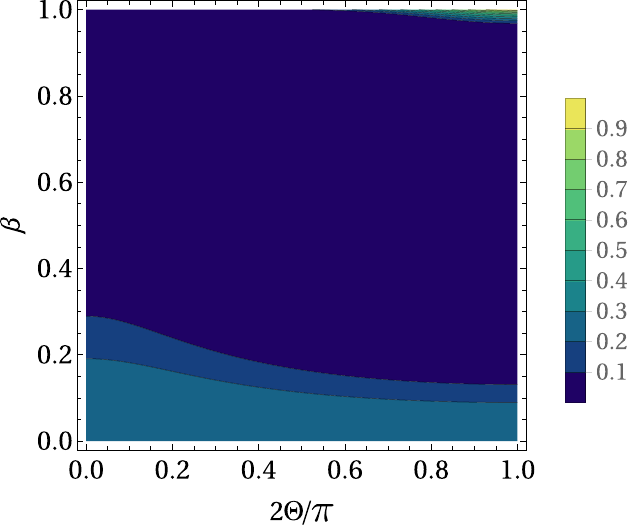}
          \caption{$w_{gg}=0.91$}
          \label{}
      \end{subfigure}
      \hspace{1cm}
      \begin{subfigure}[!]{0.3\textwidth}
          \centering
          \includegraphics[width=\textwidth]{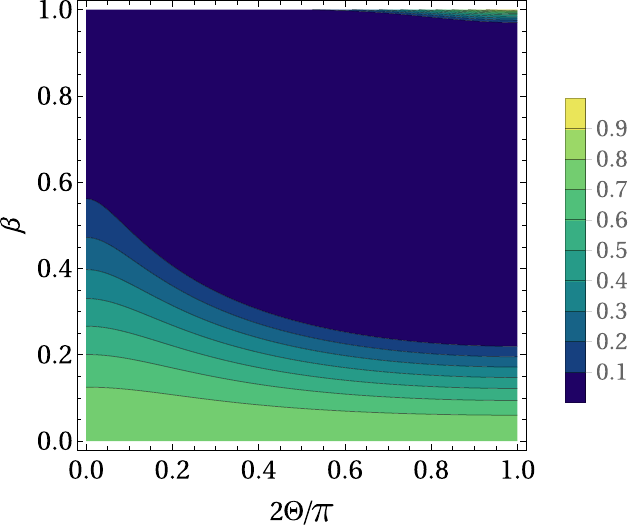}
          \caption{$w_{gg}=0.98$}
          \label{}
      \end{subfigure}
         \caption{
         Contour lines of Devetak-Winter security rate $\rateboundBB$ for entanglement-based BB84 protocol using mixed-process $\rho_{AB}^{t\bar t}$ resource states with $w_{gg}=0.91$ (left panel) and $w_{gg}=0.98$ (right panel).
         } 
         \label{fig_sec_rate_bb84_mix}
 \end{figure*}

   \begin{figure*}
      \centering
      \begin{subfigure}[!]{0.3\textwidth}
          \centering
          \includegraphics[width=\textwidth]{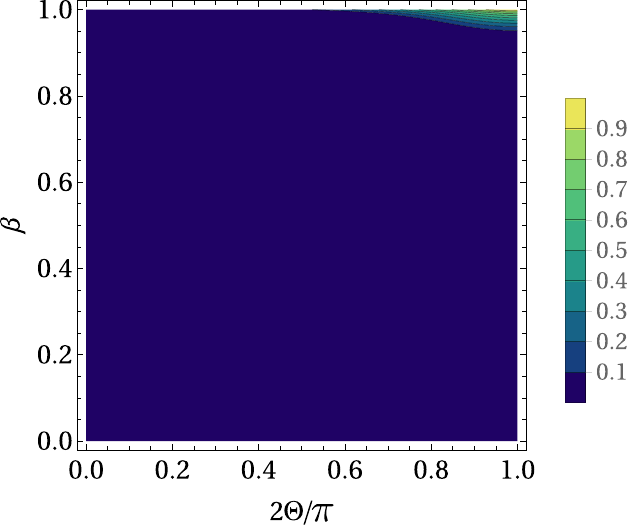}
          \caption{$w_{gg}=0.77$}
          \label{}
      \end{subfigure}
      \hfill
      \begin{subfigure}[!]{0.3\textwidth}
          \centering
          \includegraphics[width=\textwidth]{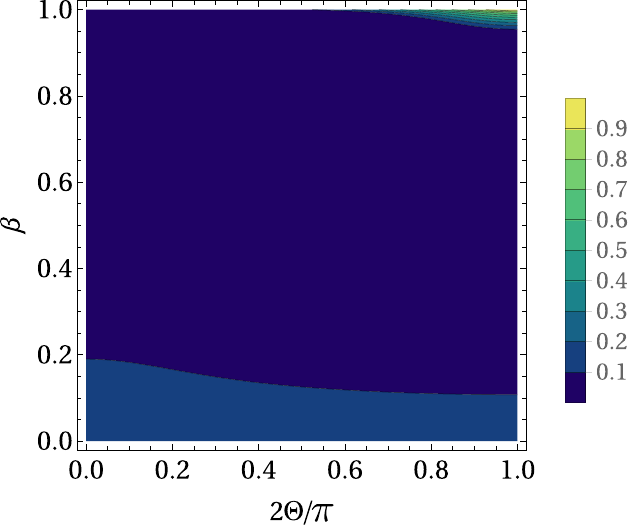}
          \caption{$w_{gg}=0.84$}
          \label{}
      \end{subfigure}
    \hfill
    \begin{subfigure}[!]{0.3\textwidth}
          \centering
          \includegraphics[width=\textwidth]{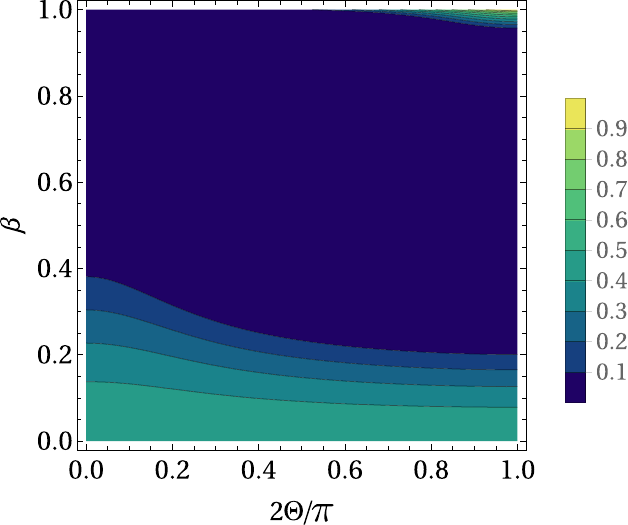}
          \caption{$w_{gg}=0.91$}
          \label{}
      \end{subfigure}
         \label{}
         \begin{subfigure}[!]{0.3\textwidth}
          \centering
          \includegraphics[width=\textwidth]{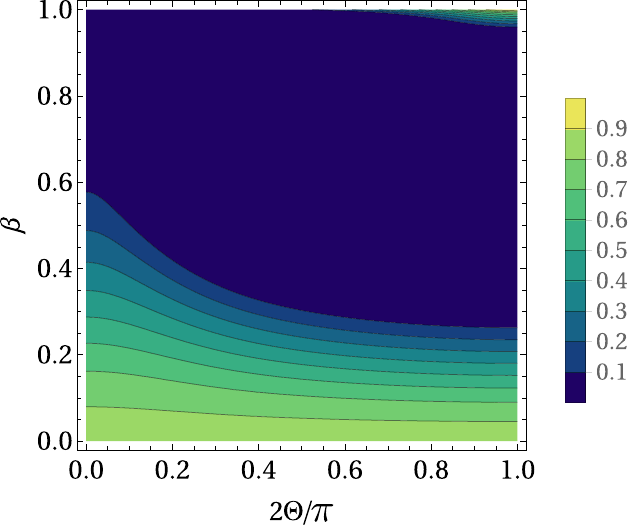}
          \caption{$w_{gg}=0.98$}
          \label{}
      \end{subfigure}
         \caption{Security rate contours $\rateboundOpt$ for the modified protocol showing enhanced performance with mixed $t\bar{t}$ processes.}
         \label{fig_sec_rate_bb84_mod_mix}
 \end{figure*}

 \begin{figure*}
      \centering
      \begin{subfigure}[!]{0.4\textwidth}
          \centering
          \includegraphics[width=\textwidth]{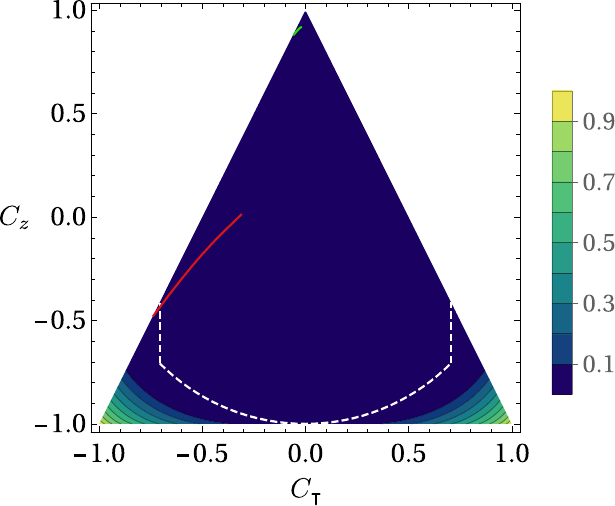}
          \caption{$\rateboundBB$ for $\rho_{\text{int}}$}
          \label{}
      \end{subfigure}
      \hspace{1cm}
      \begin{subfigure}[!]{0.4\textwidth}
          \centering
          \includegraphics[width=\textwidth]{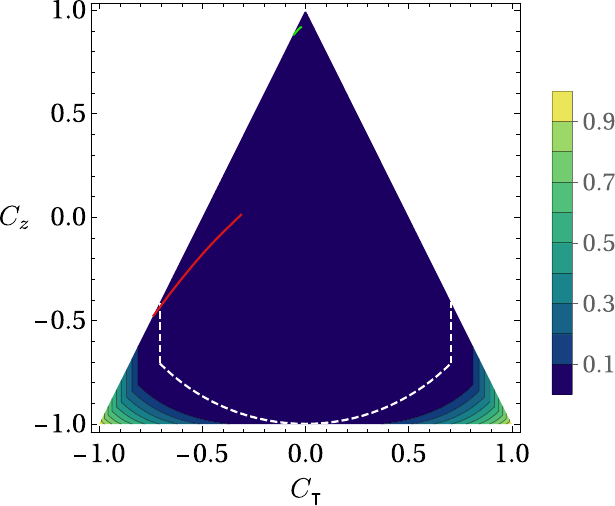}
          \caption{$\rateboundOpt$ for $\rho_{\text{int}}$}
          \label{}
      \end{subfigure}
         \caption{Contour lines of security rates $\rateboundBB$ and $\rateboundOpt$ for the integrated state $\rho_{\text{int}}$ in the $(C_\intercal, C_z)$ parameter space. The dashed white line indicates the boundary of non-local states, while red and green lines show trajectories achievable at LHC ($\sqrt{s}=13$ TeV) and Tevatron ($\sqrt{s}=2$ TeV), respectively (see Ref.~\cite{afik_quantum_2023}).} 
         \label{fig_plots_int}
 \end{figure*}

\subsubsection{Imperfect source: top-antitop pairs}


Let us examine the capabilities of top-antitop pairs produced in colliders employed as resource states of the entanglement-based quantum key distribution protocols presented in Result~\ref{res:rate_Bound}. In the first place, to explore the maximal capabilities of these systems, we will consider the states $\rho_{AB}^{t\bar t}$, introduced in Sec.~\ref{sec_top_quarks_predictability}, Eq.~\eqref{eq_top_quark_state}. Then, we will consider a set of quantum states that are experimentally reproducible in colliders, see for example Ref.~\cite{afik_quantum_2023}, which we shall denote as $\rho_{\text{int}}$.

Figure~\ref{fig_sec_rate_bb84} presents the Devetak-Winter rate $\rateboundBB$ for the entanglement-based BB84 quantum key distribution protocol utilizing the resource state $\rho_{AB}^{t\bar{t}}$. We analyze two distinct physical scenarios: the pure light quark-antiquark production process (characterized by $w_{gg}=0$) and the pure gluon-gluon production process (characterized by $w_{gg}=1$). 

In the quark-antiquark regime, non-null security rates are achieved exclusively within a region approaching the kinematic limit $(\beta,\Theta) \to (1,\pi/2)$. Conversely, the gluon-gluon production mechanism exhibits non-zero key rates $\rateboundBB > 0$ across two distinct parameter domains: a confined region near the high-energy limit $(\beta,\Theta) \to (1,\pi/2)$, and the region near the threshold production where $\beta \to 0$.

For mixed processes, $\rho_{AB}^{t\bar t}$ is generally unsuitable for quantum key distribution, except at the specific point $(\beta,\Theta)\to (1,\pi/2)$. When $w_{gg}>1/\sqrt{2}$, non-local states emerge at the threshold; however, BB84 remains ineffective at threshold until approximately $w_{gg}\approx 0.91$. Fig.~\ref{fig_sec_rate_bb84_mix} shows the contour lines of $\rateboundBB$ for $w_{gg}=0.91$ and $w_{gg}=0.98$.

The entanglement-based modified protocol characterized by Result~\ref{res:rate_Bound} enhances security for this type of resource state. In Fig.~\ref{fig_sec_rate_bb84_mod_mix}, we plot the contour lines of the bound $\rateboundOpt$ from Eq.~\eqref{eq_rate_bound_opt} for the mixed process with $w_{gg}>1/\sqrt{2}$. As we can see, for $w_{gg}=0.84$, the modified protocol enables establishing a non-zero security rate at threshold production. For values of $w_{gg}$ greater than $0.84$, we observe a substantial improvement. We choose these $w_{gg}$ values because they indicate a clear advantage of our modified protocol for mixed processes, but it is important to mention that also an improved performance is seen for pure processes, i.e. $w_{gg}=0$ and $w_{gg}=1$.

Now, while $\rho^{t\bar t}_{AB}$ is the actual spin state of top-antitop pairs naturally produced in proton-proton or proton-antiproton collisions and stands for a crucial element when studying the capabilities of such systems, it is important to mention that $\rho^{t\bar t}_{AB}$ cannot currently be directly measured or controlled~\cite{afik_quantum_2022}. The main idea at this point is that by integrating all relevant variables over a specific phase space region, the spin information can be extracted through the detection of the corresponding leptonic decay products from the top-antitop pairs~\cite{afik_quantum_2023}. However, when the spin quantization axis is event-dependent, as occurs with the diagonal basis previously employed in Sec.~\ref{sec_top_quarks_predictability}, the resulting integrated quantum state lacks physical interpretation. Thus, in the following, we will employ the fixed beam axis $\{\hat{x},\hat{y},\hat{z}\}$~\cite{atlas2024,afik_quantum_2022,afik_quantum_2023} instead of the diagonal basis employed for $\rho^{t\bar t}_{AB}$. Summarizing, this basis and at leading order perturbation theory, results in Bell-diagonal state $\rho_{\text{int}}$ for the angularly averaged state with mass integration only at threshold (i.e., between $[2m_t,M_{t\bar{t}}]$, where $2m_t$ and $M_{t\bar{t}}$ are the invariant masses before and after pair production, respectively)~\cite{afik_quantum_2023}. The correlation matrix of $\rho_{\text{int}}$ depends only on two parameters: 
\begin{align}\label{eq_corr_mat_integrated}
    \C_{\text{int}}=\text{diag}\{c_1=C_\intercal,c_2=C_\intercal,c_3=C_z\},
\end{align}
which are ultimately determined by the mass threshold integration limit $M_{t\bar{t}}$, which depends on the collision energy $\sqrt{s}$.

Let us now examine the security rates for the integrated state $\rho_{\text{int}}$. Fig.~\ref{fig_plots_int} shows the contour lines of $\rateboundBB$ and $\rateboundOpt$ for values of $C_\intercal$ and $C_z$ that define a positive semi-definite $\rho_{\text{int}}$. The dashed white line indicates non-local states, while the red and green lines represent trajectories achievable at the LHC ($\sqrt{s}=13$ TeV) and Tevatron ($\sqrt{s}=2$ TeV), respectively~\cite{afik_quantum_2023}. 

As we can see, the states that can be currently prepared at the LHC yield a null security rate for either BB84 or our modified protocol. 
\section{Concluding remarks}

In this work, we investigated the transmission of classical information by quantum means from the point of view of statistical learning theory. Previous approaches have quantified this process through measures such as the Shannon entropy, leading to the notion of accessible information~\cite[Ch.20]{wilde2017}, and through classical–quantum entropic uncertainty relations~\cite{coles2017}, where the relevant figure of merit is the probability of error equivalent to the quantum bit error rate studied here.

Building on this context, we focused on two complementary predictability measures: the Bayes risk $\BayesRisk$ [Eq.\eqref{eq:BayesRisk}] and the inference variance $\inferenceVariance$ [Eq.\eqref{eq_inference_variance_def}]. We derived analytical optimizations of both measures for arbitrary two-qubit resource states (see Results~\ref{res:Bayesrisk} and~\ref{res:QuadratEntropy}), and established a clear operational link between enhanced predictability, quantified by the averaged minimal Bayes risk and inference variance for all $A$-observables, and Einstein–Podolsky–Rosen steering in Bell-diagonal states: only steerable states surpass the local predictability thresholds set by uncorrelated measurements.

We further explored the robustness and limitations of quantum correlations in realistic scenarios, including Bell pairs subject to local noise and the spin-density matrices of top–antitop quark pairs. One practical implication is the enhancement of entanglement-based quantum key distribution: By selecting Bob's measurement according to the minimal Bayes risk, one can achieve higher secure-key rates than the conventional BB84 protocol, maintaining security even under stronger noise conditions.

Our analysis also provides a novel study into the potentialities of quantum states produced in colliders for quantum cryptography. In particular, our investigation of top--antitop quantum states encompassed both the theoretical limits imposed by non-averaged states $\rho^{t\bar t}_{AB}$, which establish upper bounds on quantum information capacity, and the experimentally accessible integrated states $\rho_{\text{int}}$ realizable within current LHC measurement frameworks. Despite the null Devetak--Winter rates observed under present conditions (Fig.~\ref{fig_plots_int}), the landscape is rapidly evolving: recent proposals from the CERN collaboration~\cite{afik2025} demonstrate viable pathways for advancing correlation measurement techniques, potentially unlocking the preparation of entangled states suitable for quantum cryptographic applications in high-energy physics environments.

Overall, our work connects foundational information-theoretic measures with operational quantum communication tasks, examines applications to high-energy physics platforms, and identifies pathways for advancing the use of quantum correlations in secure communication.

\section*{Acknowledgments}
The authors thank Mateus Araújo for his valuable assistance and insightful suggestions regarding Quantum Key Distribution. This research was supported by the Q-CAYLE project, funded by the European Union-Next Generation UE/MCIU/Plan de Recuperacion, Transformacion y Resiliencia/Junta de Castilla y Leon (PRTRC17.11), and also by RED2022-134301-T and PID2023-148409NB-I00, financed by MI-CIU/AEI/10.13039/501100011033.  
The financial support of the Department of Education of the Junta de Castilla y León and FEDER Funds is also gratefully acknowledged (Reference: CLU-2023-1-05). 
\bibliography{library.bib}
\onecolumngrid
\appendix

\section{Bayes Risk optimized} \label{app:BayesRiskOpt}

In this Section, we will derive Result~\ref{res:Bayesrisk}.

An arbitrary two-qubit quantum state can be written as:
\begin{align}
        \rho_{AB}\!=\!\frac{1}{4}\!\left(\!\mathbbm{1}_{4}\!+\!\vec{t}_{A}\!\cdot \!\vec{\sigma}_A\!\otimes\!\mathbbm{1}\!+\!\mathbbm{1}\otimes \vec{t}_{B}\!\cdot\!\vec{\sigma}_B\!+\!\sum_{i,j=1}^{3}\!C_{i j} \, \!\sigma_A^{i} \otimes \sigma_B^j\right),\label{eq_Fanoform}
\end{align} 
where $\vec{t}_{A}\in\BlochBall$ and $\vec{t}_{B}\in\BlochBall$ are the Bloch vectors ($\BlochBall$ stands for the tridimensional real unit Ball) of systems $A$ and $B$, Eqs.~\eqref{eq_Fanoform_tA} and~\eqref{eq_Fanoform_tB} respectively, $\vec{\sigma}$ are the Pauli matrices, and $C_{ij}$ corresponds to the elements of the correlation matrix $\textbf{C}$ [Eq.~\eqref{eq_Fanoform_C}].

The corresponding local observable operators of each subsystem can be expressed as $\mathcal{O}(\vec{a})=\vec{a}\cdot\vec{\sigma}_A$ and $\mathcal{O}(\vec{b})=\vec{b}\cdot\vec{\sigma}_B$, defining the following measurement operators:
\begin{align}
 M_x(\vec{a})&=\frac{1}{2}[\mathbbm{1}+(-1)^{x}\vec{a}\cdot\vec{\sigma}], \label{eq_projectorsA} \\ M_y(\vec{b})&=\frac{1}{2}[\mathbbm{1}+(-1)^{y}\vec{b}\cdot\vec{\sigma}], \label{eq_projectorsB}
\end{align}
where $\va,\vb\in \BlochSphere$.

If for example Bob measures in his subsystem $\mathcal{O}(\vb)$ over the joint state $\rho_{AB}$, and obtains the result $Y=y$, the conditional states occupying Alice's system result in:
\begin{align}
    \rho_{A|y}&=\frac{\text{Tr}_B\left[ \mathbbm{1}\otimes M_y(\vb) \rho_{AB} \right]}{P(y)}=\frac{1}{2}(\mathbbm{1}+\vec{t}_{A|y}\cdot\vec{\sigma}), \label{eq_conditional_states}\\
    P(y)&=\text{Tr}M_y(\vb)\rho_B=\frac{1}{2}[1+(-1)^y\vb\cdot\tB],\\
    \vec{t}_{A|y}&=\frac{\tA+(-1)^y \C\vb}{2P(y)}.\label{eq_conditional_bloch_vect}
\end{align}

If Alice also takes a measurement of $\mathcal{O}(\va)$, the joint probability distribution of obtaining results $X$ and $Y$ is,
\begin{align}
P(x,y)&=\Tr{M_x(\vec{a})\otimes M_y(\vec{b}) \ \rho_{AB}}=P(y)P(x|y)\nonumber\\&=\frac{1}{2}\left\{P(y)+(-1)^x \vec{a}\cdot\frac{[\vec{t}_A+(-1)^y\textbf{C}\vec{b}]}{2}\right\},\label{eq:JoinProb2}   
\end{align}
and the conditional expectation $\ConditionalExp$, see Eq.~\eqref{eq:CondExp}, reads
\begin{align}\label{eq:CondExp2}
    \ConditionalExp(y)=\frac{1}{2}\left\{1-\vec{a}\cdot\left[\frac{\vec{t}_A+(-1)^y (\textbf{C}\vec{b})}{2P(y)}\right]\right\}.
\end{align}

Let us use now the expression Eq.~\eqref{eq:BayesRisk}:
\begin{align}
    \BayesRisk = \frac{1}{2}-\frac{1}{2}\average{|2\ConditionalExp(y)-1|}.
\end{align}
The conditional expectation satisfies:
\begin{align}
    \ConditionalExp(y)>1/2 \ \iff \ -\vec{a}\cdot\vec{t}_A>(-1)^y \vec{a}\cdot \C \vec{b}.
\end{align}
It follows:
\begin{align}
    \ConditionalExp(0)&>1/2 \ \And \ \ConditionalExp(1)>1/2 \ \iff \ \va\cdot\tA<\va\cdot\C\vb<-\va\cdot\tA, \label{eq:CondBayes1} \\
     \ConditionalExp(0)&<1/2 \ \And \ \ConditionalExp(1)<1/2 \ \iff \ -\va\cdot\tA<\va\cdot\C\vb<\va\cdot\tA, \\
     \ConditionalExp(0)&<1/2 \ \And \ \ConditionalExp(1)>1/2 \ \iff \ -\va\cdot\C\vb<\va\cdot\tA<\va\cdot\C\vb, \\
     \ConditionalExp(0)&>1/2 \ \And \ \ConditionalExp(1)<1/2 \ \iff \ \va\cdot\C\vb<\va\cdot\tA<-\va\cdot\C\vb.\label{eq:CondBayes4}
\end{align}
Thus,
\begin{align}
    \BayesRisk=\begin{cases}
       \frac{1}{2}(1-|\va\cdot\tA|) & -|\va\cdot\tA|<\va\cdot\C\vb<|\va\cdot\tA| \\ 
        \frac{1}{2}(1-|\va\cdot\C\vb|) & -|\va\cdot\C\vb|<\va\cdot\tA<|\va\cdot\C\vb|.
    \end{cases}
\end{align}
If $|\va\cdot\tA|>|\C^\intercal\va|\implies$ there is no $\vb$ such that $-|\va\cdot\C\vb|<\va\cdot\tA<|\va\cdot\C\vb|$; thus, $\BayesRisk$ becomes independent of $\vb$ and the decision $\BayesClass$ is determined by the sign of $\va\cdot\tA$, see Eqs.~\eqref{eq:CondBayes1}-\eqref{eq:CondBayes4}. Otherwise, $|\va\cdot\tA|<|\C^\intercal\va|$, we can minimize the Bayes risk by taking $\boboptimalMeasurement=\C^\intercal\va/|\C^\intercal\va|$; Finally:
\begin{align}
    \BayesRiskMin=\min_{\vb\in\BlochSphere}\BayesRisk=\begin{cases}
       \frac{1}{2}(1-|\va\cdot\tA|) & |\C^\intercal\va|<|\va\cdot\tA| \\ 
        \frac{1}{2}(1-|\C^\intercal\va|) & \text{otherwise}.
    \end{cases}
\end{align}


\section{Inference variance minimization}\label{app:proof R1}

The inference variance $\inferenceVariance$, Eq.~\eqref{eq_inference_variance_def}, is
\begin{align}
    \inferenceVariance =\sum_{xy}[x-\ConditionalExp(y)]^{2} P(x,y),
\end{align}
where $\ConditionalExp$ is the conditional expectation, for the corresponding joint probability distribution $P(x,y)$, see Eqs.~\eqref{eq:CondExp2} and \eqref{eq:JoinProb2}, respectively. By doing the required algebra, we have that:
\begin{align}
    \inferenceVariance &= \frac{1}{4}[1-\certainty(\vb)], \\
    \certainty(\vb) &= \frac{(\va\cdot\tA)^2+(\va\cdot \C \vb)^2-2(\tB\cdot\vb)(\va\cdot\tA)(\va\cdot \C \vb)}{1-(\tB\cdot\vb)^2}.
\end{align}

We have to find the maximum thus of the following quantity:
\begin{equation}
\max_{\vb\in\BlochSphere} \ \certainty(\vb) :=\max_{\vb\in\BlochSphere}\frac{(\va\cdot\tA)^2+(\va\cdot \C \vb)^2-2(\tB\cdot\vb)(\va\cdot\tA)(\va\cdot \C \vb)}{1-(\tB\cdot\vb)^2},
\end{equation}
constrained by $\mathcal{G}(\vb)=|\vb|^{2}-1=0$. 

By following the Lagrange multipliers method, the critical points of $\certainty(\vb)$, constrained by $\mathcal{G}(\vb)$, satisfy:
\begin{equation}
\frac{\partial \certainty(\vb)}{\partial b_{i}}=\lambda\frac{\partial \mathcal{G}(\vb)}{\partial b_{i}}, \qquad i= 1,2,3. \label{eq_Lagrange}
\end{equation}
Doing the required calculations, and using $|\vb|^{2}=1$, the Lagrange multiplier $\lambda$ can be written as:
\begin{equation}
\lambda=\left(\frac{(\va\cdot \C\vb)-(\va\cdot\tA)(\tB\cdot\vb)}{1-(\tB\cdot \vb)^2}\right)^2.
\end{equation}
Let us denote the optimal direction $\boptQuadraticEntropy$ as the unit vector $\vb_\ast/|\vb_\ast|$. The substitution of $\lambda$ into Eq.~\eqref{eq_Lagrange} leads to:
\begin{align}
\alpha(\vb_{\ast}) (\textbf{C}^\intercal\va)+\beta(\vb_{\ast})\tB = \gamma(\vb_{\ast}) \vb_{\ast}, \label{eq_comlineal}
\end{align}
where the auxiliary coefficients are defined as follows:
\begin{align}\label{eq_coef}
& \alpha(\vb_\ast)=\left[(\va\cdot \C\vb_\ast)-(\va\cdot\tA)(\tB\cdot\vb_\ast)\right] \left[|\vb_\ast|^2-(\tB\cdot\vb_\ast)^2\right],\nonumber \\
& \beta(\vb_\ast)=(\tB\cdot\vb_\ast) \left[(\va\cdot\tA)^2 |\vb_\ast|^2+(\va\cdot \C\vb_\ast)^2\right]-(\va\cdot \C\vb_\ast)(\va\cdot\tA)\left[|\vb_\ast|^2+(\tB\cdot\vb_\ast)^2\right], \\
&\gamma(\vb_\ast)=\left[(\va\cdot \C\vb_\ast)-(\va\cdot\tA)(\tB\cdot\vb_\ast)\right]^{2}. \nonumber
\end{align}
From Eq.~\eqref{eq_comlineal}, we can see that the vector $\vb_{\ast}$ is a linear combination of $\C^\intercal\va$ and $\tB$:
\begin{equation}
\vb_\ast=d_{1}\C^\intercal\va+d_{2}\tB,
\end{equation}
being:
\begin{align}
& d_{1}=\alpha(\vb_\ast)/ \gamma(\vb_\ast), \nonumber \
& d_{2}=\beta(\vb_\ast)/ \gamma(\vb_\ast).
\end{align}
Therefore, by using the explicit expressions of the coefficients $\alpha$, $\beta$ and $\gamma$ in Eq.~\eqref{eq_coef}, we can show that:
\begin{equation}
d_{2}=d_{1}\frac{(\va\cdot \C\tB-\va\cdot\tA)}{(1-|\tB|^{2})},
\end{equation}
thus, the optimal unit vector $\boptQuadraticEntropy=\vb_\ast/|\vb_\ast|$ becomes:
\begin{equation}
\boptQuadraticEntropy=\pm\frac{(1-|\tB|^{2})\C^\intercal\va+(\va\cdot \C\tB-\va\cdot\tA)\tB}{|(1-|\tB|^{2})\C^\intercal\va+(\va\cdot \C\tB-\va\cdot\tA)\tB|}.
\end{equation}
This critical point leads to
\begin{align}
    \certainty(\boptQuadraticEntropy)=\certaintyOpt=\frac{|\va\cdot \C \tB-\va\cdot\tA|^2+|\C^\intercal\va|^2(1-|\tB|^{2})}{1-|\tB|^{2}}.
\end{align}
Let us see that $\certaintyOpt$ is indeed the maximum. Instead of calculating the Hessian matrix, we can show directly that:
\begin{align}
    \certainty(\vb)\leq\frac{|\va\cdot \C \tB-\va\cdot\tA|^2+|\C^\intercal\va|^2(1-|\tB|^{2})}{1-|\tB|^{2}}, \ \forall\vb.
\end{align}
Any three-dimensional vector can be written as:
\begin{align}
    \vb = \alpha_1\C^\intercal\va+\alpha_2\vec{t}_B+\alpha_3(\C^\intercal\va\times\vb).
\end{align}
Thus, it holds $\certainty(\vb)=\certainty(\alpha_1\C^\intercal\va+\alpha_2\vec{t}_B)$, and by doing the required calculations, we can see that the condition $\certainty(\alpha_1\C^\intercal\va+\alpha_2\vec{t}_B)\leq \certaintyOpt $ is equivalent to:
\begin{align}
    \frac{[(\C^\intercal\va)^2t_B^2-(\va\cdot \C\vec{t}_B)^2][\alpha_1a\cdot\C\tB-\alpha_1\va\cdot\tA-\alpha_2(1-t_B^2)]^2}{(1-t_B^2)\left\{[(\va\cdot\C\tB)^2-(\C^\intercal\va)^2]\alpha_1^2-2\alpha_1\alpha_2(\va\cdot\C\tB)(1-t_B^2)-t_B^2(1-t_B^2)\alpha_2^2\right\}} \leq 0
\end{align}
Because $(\C^\intercal\va)^2t_B^2-(\va\cdot \C\vec{t}_B)^2=(\C^\intercal\va)^2t_B^2-(\C^\intercal \va\cdot \vec{t}_B)^2\geq 0$ and $t_B^2\leq 1$, we have just to demonstrate:
\begin{align}
    &[(\va\cdot\C\tB)^2-(\C^\intercal\va)^2]\alpha_1^2-2\alpha_1\alpha_2(\va\cdot\C\tB)(1-t_B^2)-t_B^2(1-t_B^2)\alpha_2^2\leq 0 \ \iff\\
    &-|\alpha_1 \C^\intercal\va +\alpha_2\tB|^2+[\alpha_1 \va\cdot\C\tB+\alpha_2t_B^2]^2\leq 0 \label{eq:lastineq}
\end{align}
This last inequality holds because $\alpha_1 \va\cdot\C\tB+\alpha_2t_B^2=(\alpha_1 \C^\intercal\va+\alpha_2\tB)\cdot\tB$ and therefore, defining $\vec{\tilde b} = \alpha_1 \C^\intercal+\alpha_2\tB$, Eq.~\eqref{eq:lastineq} is equivalent to $|\vec{\tilde b}|^2\geq (\vec{\tilde b}\cdot \tB)^2$.

Consequently, the minimized inference variance $\inferenceVariance$ turns out to be:
 \begin{align}
        \inferenceVarianceMin &= \min_{\vb\in\BlochSphere}\inferenceVariance = \frac{1}{4}\left(1-\certaintyOpt\right), \\
        \certaintyOpt &= \frac{|\va\cdot \C \tB-\va\cdot\tA|^2+|\C^\intercal\va|^2(1-|\tB|^{2})}{1-|\tB|^{2}}.  
    \end{align}


\section{Obtaining the \predictability}\label{app:AveragePredictability}
\subsection{Average Minimal Bayes risk}\label{app:AverageBayesRisk}
\textbf{Correlation scenario.} The average of the minimal Bayes risk $\BayesRiskMin$ over all directions of $\va \in \BlochSphere$, when the correlations of the subsystems are taken into account, Eq.~\eqref{eq_assumption}, is computed using the following surface integral:
\begin{align}
    \overline{\BayesRiskMin}(\tA, \tB, \C) &= \frac{1}{4\pi} \iint_{S=\BlochSphere}\frac{1}{2}\left(1-|\C^\intercal\va|\right) \, \mathrm{d}S, \nonumber \\
    &= \frac{1}{4\pi} \iint_{\BlochSphere}\frac{1}{2}\left(1-\sqrt{\va \cdot(\C \C^\intercal \va)}\right) \, \mathrm{d}S.
\end{align}
It is worth noting that the presence of the square root makes this integral somewhat complex to compute. However, using the \textit{singular value decomposition} (SVD), we can write $\C=\textbf{R}_1\C_{d}\textbf{R}_2$, where $\textbf{R}_i$ are rotation matrices, and $\C_{d}=\text{diag}\{ c_1, c_2, c_3\}$. If follows,
\begin{align*}
    \va \cdot(\C \C^\intercal \va) =\va' \cdot \C_d^2 \va',
\end{align*}
where $\va'=\textbf{R}_1 \va$. Because of the rotational symmetry of the integral with respect to $\va$, we can just simply integrate over $\va$, and take $\C=\C_d$.

Expressing $\va^\intercal$ in spherical coordinates $(\sin \theta \cos \varphi, \sin \theta \sin \varphi, \cos \theta)$, the average of $\BayesRiskMin$ becomes:
\begin{align}
    \overline{\BayesRiskMin} = \frac{1}{2}\left(1-\frac{1}{4\pi} \int_{0}^{\pi} \mathrm{d}\theta\int_{0}^{2\pi}\mathrm{d} \varphi \sqrt{g}\right),
\end{align}
where $g:=c_{3}^{2}\cos^{2}\theta \sin^{2}\theta+(c_{2}^{2}\sin^{2} \varphi+c_{1}^{2} \cos^{2} \varphi)\sin^{4} \theta$. This integral is determined by the surface area of an ellipsoid, for example defined by:
\begin{equation}
    \frac{x^{2}}{a^{2}}+\frac{y^{2}}{b^{2}}+\frac{z^{2}}{c^{2}}=1.
\end{equation}
Thus, by establishing a correspondence between the semi-axes of the ellipsoid $\{ a,b,c\}$ and the diagonal elements of the correlation matrix $\{ c_1,c_2,c_3\}$ as follows: 
\begin{align*}
    |c_{1}| \rightarrow cb, \quad
    |c_{2}|\rightarrow ca, \quad
    |c_{3}|\rightarrow ab,
\end{align*}
the solution of the integral is found to be:
    \begin{equation}
        \frac{1}{4\pi} \int_{0}^{\pi} \mathrm{d}\theta\int_{0}^{2\pi}\mathrm{d} \varphi \sqrt{g}= |c_{1}|R_{G}\left(\frac{|c_{2}|^{2}}{|c_{1}|^{2}},\frac{|c_{3}|^{2}}{|c_{1}|^{2}},1 \right),
    \end{equation}
with $R_{G}$ a Carlson symmetric elliptic integral. Consequently, the average of the minimal Bayes risk $\BayesRiskMin$ is given by:
\begin{equation}\label{eq:BayesRiskAverage1app}
\overline{\BayesRiskMin}(\C) = \frac{1}{2}\left[1-|c_{1}|R_{G}\left(\frac{|c_{2}|^{2}}{|c_{1}|^{2}},\frac{|c_{3}|^{2}}{|c_{1}|^{2}},1\right)\right].
\end{equation}
\\

\textbf{Local Information scenario.} When subsystems are uncorrelated, meaning only local information from subsystem $A$ is considered, the average minimal Bayes risk across all directions of $\va$ is calculated through:
\begin{align}
    \Bayesriskaveragelocal(\tA) = \frac{1}{4\pi} \iint_{S=\BlochSphere}\frac{1}{2}\left(1-|\va \cdot \tA|\right) \, \mathrm{d}S.
\end{align}
To solve this integral, we can use the above result in Eq.~\eqref{eq:BayesRiskAverage1app} for the correlation scenario. First of all, note that if we choose $\C'= \frac{\tA \tA^\intercal}{|\tA|}$, $|\C'^\intercal\va|= |\va \cdot\tA|$ for all $\va$.

Furthermore, we can diagonalize the correlation matrix $\C'$ by means:
\begin{equation}
    \textbf{O}^{-1} \C' \textbf{O}=\C'_{d},
\end{equation}
such that, $\textbf{O} \in \mathbbm{R}^{3\times3}$ is an invertible matrix composed of the eigenvectors of $\C'$. This diagonalization yields $\C'_{d} = \textrm{diag}\{|\tA|, 0, 0\}$. Consequently, we have:
\begin{equation}
    \iint_{\BlochSphere}|\va \cdot \tA| \mathrm{d}S=|\tA| R_{G}\left(0,0,1 \right)= \frac{|\tA|}{2}.
\end{equation}
Therefore, the average minimal Bayes risk for this uncorrelated scenario is:
\begin{align}\label{eq:BayesRiskAverage2app}
    \BayesRiskMinLocal(\tA) = \frac{1}{2}\left(1-\frac{|\tA|}{2}\right).
\end{align}

\subsection{Average minimal inference variance}\label{app:AverageQuadraticEntropy}
The average of the minimal inference variance $\inferenceVarianceMin$ in Eq.~\eqref{eq_quadraticEnt_opt} over all the directions of the vector $\va \in \BlochSphere$ is computed as follows:
\begin{align}
    \inferenceVarianceMinAverage(\tA, \tB, \C)=\frac{1}{4\pi} \iint_{S=\BlochSphere}\frac{1}{4}\left(1-\certainty\right) \, \mathrm{d}S,
\end{align}
where $\certainty$ is given by:
 \begin{align}
        \certainty(\va,\tA, \tB, \C) &= \frac{|\va\cdot (\C \tB-\tA)|^2+|\C^\intercal\va|^2(1-|\tB|^{2})}{1-|\tB|^{2}}.
    \end{align}
By defining $\vec{x}:= \C \tB-\tA$, it is possible to express $|\va \cdot \vec{x}|^{2}+(1-|\tB|^{2})|\C^\intercal \va|^{2}=\va \cdot [\vec{x}\vec{x}^\intercal+(1-|\tB|^{2})\C \C^\intercal]\, \va$. Hence, the average of the inference variance can be rewritten as 
\begin{align}
    \inferenceVarianceMinAverage(\tA, \tB, \C)=& \frac{1}{4\pi} \iint_{\BlochSphere}\frac{1}{2}\left(1-\va \cdot \textbf{M} \ \va\right) \, \mathrm{d}S,
\end{align}
where $\textbf{M}$ is a $3 \times 3$ matrix defined by:
\begin{equation}
\textbf{M}:= \frac{(\C \tB-\tA)(\C \tB-\tA)^\intercal+(1-|\tB|^{2})\C \C^\intercal}{1-|\tB|^{2}}.
\end{equation}
Finally, by applying the following result:
\begin{equation}\label{eq:Resultrace}
    \frac{1}{4\pi}\iint_{S} \va \cdot \textbf{M}\va \, \textrm{d}S = \frac{1}{3} \text{Tr}[\textbf{M}],
\end{equation} 
it is concluded that the average of the minimal inference variance is expressed as:
\begin{align}
     \inferenceVarianceMinAverage(\tA, \tB, \C) =& \frac{1}{4}\left(1-\frac{1}{3} \text{Tr}[\textbf{M}]\right) \nonumber \\
     =& \frac{1}{4}\left(1-\frac{\|\C\tB-\tA\|^2+(1-|\tB|^{2})\|\C\|^2}{3(1-|\tB|^{2})}\right),
\end{align}
with $\|\textbf{A} \|^{2}:= \text{Tr}[\textbf{A}\textbf{A}^{\dagger}]$ being the Hilbert-Schmidt inner product. 
\\

Furthermore, the average of the minimal quadratic entropy when only local information is considered is determined by:
\begin{align}
\inferenceVarianceMinLocalAverage(\tA) &= \frac{1}{4\pi} \iint_{S=\BlochSphere}\frac{1}{4}\left(1-|\va \cdot \tA|^{2}\right) \mathrm{d}S \\
&= \frac{1}{4\pi} \iint_{\BlochSphere}\frac{1}{4}\left(1-\va \cdot \textbf{N} \va\right) \mathrm{d}S,
\end{align}
where $\textbf{N}: = \tA \tA^\intercal$. Thus, applying the result in Eq.~\eqref{eq:Resultrace}, we obtain:
\begin{equation}
\inferenceVarianceMinLocalAverage(\tA) = \frac{1}{2}\left(1-\frac{|\tA |^{2}}{3} \right).
\end{equation}

\section{Classical-quantum states cannot improve the local \predictability threshold as measured by $\inferenceVarianceMinAverage$} \label{app_improving_LU_quadratic_entr}
Let us consider a set of three inequalities, known as \textit{positivity conditions} \cite{GAMEL2016}, that establish the positivity requirement for a quantum state, which are defined as follows:
\begin{align}\label{eq_posi_cond}
     &4-\| \textbf{r} \|^{2} \geq 0, \nonumber \\
     &( \tA^{\dagger} \, \C \, \tB- \text{det} \, \C)-(\| \textbf{r} \|^{2}-2) \geq 0,  \nonumber \\
     &8( \tA^{\dagger}  \, \C \, \tB- \text{det} \, \C)+(\| r \|^{2}-2)+8\tA^{\dagger} \, \C \, \tB -4(\| \tA \|^{2}\| \tB \|^{2}+\|\tA^{\dagger} \, \C \|^{2}+\| \C \, \tB \|^{2}+\| {\tilde{\C}}\|^{2})\geq 0, 
\end{align}
with $ \| \textbf{r} \|^{2}:= 1+\| \tA \|^{2}+\| \tB \|^{2}+\| \C \|^{2}$ and $\tilde{\C}$ the cofactor matrix of ${\C}$. Furthermore, this set of inequalities allows us to visualize and parametrize the state space. Consequently, we define the three-dimensional region, $\mathcal{T}$, defined by the diagonal elements of the correlation matrix $\C=\text{diag}\{c_{1}, c_{2}, c_{3}\}$ subject to the positivity conditions given in Eq.~\eqref{eq_posi_cond}. Thus, this region $\mathcal{T}$ constitutes a tetrahedron, representing the state space of all well-defined bipartite quantum states  for arbitrary Bloch vectors $\tA$ and $\tB$.

Let $\rho_{A B}^{cq}$ be the density operator corresponding to a \textbf{classical-quantum states}, defined as follows:
\begin{equation}\label{eq_cq_state}
\rho_{A B}^{cq}=\sum_{x \in \mathcal{X}} P(x) M_{x}(\hat{n}_{A}) \otimes \rho_{B}^{x}(\tB^{x})
\end{equation}
where
\begin{gather*}
\tA=\sum _{x} P(x) (-1)^{x} \hat{n}_{A}, \\
 \tB=\sum _{x} P(x) \hat{t}_{B}^{x}, \\
 C_{ij}= (\hat{n}_{A})^{i} \sum_{x} P(x) (-1)^{x}(\tB^{x})^{j}, \quad i,j=1,2,3.
\end{gather*}
Note that, the correlation matrix of the classical-quantum states $\textbf{C}$ can be expressed as $\textbf{C}=\vec{u} \, \vec{v}^\intercal$, where $\vec{u}:= \hat{n}_{A}$ and $\vec{v}:= \sum_{x} P(x) (-1)^{x} \, \vec{t}_{B}^{x}$. Consequently, the $3\times3$ matrix $\textbf{C}$ has rank one, meaning all its columns (or rows) are linearly dependent, resulting in a null determinant, $\text{det} \, \C=0$. 

Applying the positivity conditions from Eq.~\eqref{eq_posi_cond}, we therefore find that:
\begin{equation}
    \text{det}\textbf{C} = 0 \Rightarrow \inferenceVarianceMin  \geq 1/6.
\end{equation}
Therefore, the classical-quantum states cannot improve the local \predictability threshold as measured by $\quadraticentropyaverage$.


\section{Steering inequalities and ellipsoid} \label{app_steering}

Consider a quantum communication scenario between two parties, Alice and Bob, who wish to share quantum correlations. Bob operates the source, preparing and transmitting quantum states to Alice. 

An entirely classical source generates \textit{local hidden states}, specifically, Bob sends $\rho_k$ to Alice with probability $p_k$. The joint state, in this case, results to be a quantum-classical state of the form:
\begin{align}\label{eq_classical_quantum_lhs}
    \rho_{\text{LHS}}=\sum_k p_k \rho_{k} \otimes M_k,
\end{align}
where $\{M_k\}$ represents a set of orthonormal projectors.
Then, if Alice and Bob take local generalized measurements $M_A=\{M_{A,x}\}_x$ and $M_B=\{M_{B,y}\}_y$, the joint probability distribution of getting results $x$ and $y$ is:
\begin{align}\label{eq_lhs_measurements}
    P_{\text{LHS}}(x,y)=\sum_k p_k \Tr{M_{A,x}\rho_k} p(y|M_B,k)
\end{align}
being $p(y|M_B,k)=\Tr{M_{B,y} M_k}$, any possible probability distribution that a classical source may produce.
Now, how could Alice and Bob agree that Bob's source can generate actual quantum-correlated joint states? A simple answer is that they have to demonstrate that the joint state does not have the form in Eq.~\eqref{eq_classical_quantum_lhs}. One way to do this is by constructing inequalities based on taking local measurements, described by Eq.~\eqref{eq_lhs_measurements}.
A seminal work in this regard is Ref.~~\cite{cavalcanti2009}, in which it is considered the measurement of $n$ different sharp observables $\{\sharpObs{X}^i\}_{i=1}^n$, over each local system $X\in\{A,B\}$. The main result, for a two-qubit system, can be written as follows~\cite{cavalcanti2009}: Any local hidden state $\rho_{\text{LHS}}$, Eq.~\eqref{eq_classical_quantum_lhs}, holds
\begin{align}
    F_n^{\text{CJWR}}(\rho,\sharpObs{})=\frac{1}{\sqrt{n}}\left|\sum_{i=1}^n\left<\sharpObs{A}^i\otimes \sharpObs{B}^i\right>_\rho\right|\leq 1,
\end{align}
for a set of measurements $\sharpObs{}=\{\sharpObs{A}^i\otimes\sharpObs{B}^i\}_i$, such that $\sharpObs{A}^i$ are maximally incompatible measurements (i.e. any unitary transformation of the three Pauli operators). Therefore, if there exists a set of measurements over a state $\rho,$ for which $F_n^{\text{CJWR}}(\rho,\sharpObs{})>1$, it follows that $\rho$ is \textit{steerable}.
Particularly, an important set of observables is $\sharpObs{}'=\{\mathcal{O}^i\otimes\mathcal{O}^i\}_{i=1}^3$ being $\mathcal{O}^i=\vec n_i\cdot\vec\sigma$ with directions $\{\vec n_i\}_i$ fixed by the eigenbasis of the correlation matrix of $\rho$ in the Fano form, see Eq.~\eqref{eq_Fanoform_C}~\cite{costa2016}. From the observables $\sharpObs{}'$, two inequalities can be thus defined: If Alice and Bob take two measurements ($n=2$) of the three observables in $\sharpObs{}'$, we have
\begin{align}\label{eq_steering_ineq_two_meas}
    F_2^{\text{CJWR}}(\rho,\sharpObs{}')=\sqrt{\C_d^2-\C_{d,\min}^2}\leq 1.
\end{align}
If three measurements are taken,
\begin{align}\label{eq_steering_ineq_three_meas}
    F_3^{\text{CJWR}}(\rho,\sharpObs{}')=\sqrt{\C_d^2}\leq 1.
\end{align}

Additionally, for Bell-diagonal (or $T$) states, there exists a criterion based on all local observables, namely, a Bell-diagonal state $\rho_{\text{BD}}$ is \textit{steerable} if~\cite{jevtic2015} 
\begin{align}\label{eq_steering_ineq_haar_meas}
    F_\text{Haar}(\rho)=\int d\Omega\sqrt{\vec n \cdot\C \C^\intercal \vec n}>2\pi,
\end{align}
being $\vec{n}$ an unit vector and $d\Omega$ the differential solid angle in the unit sphere. Note that in Appendix~\ref{app:AverageBayesRisk} we showed that the results of the previous integration can be written in terms of the Carlson symmetric elliptic integral, $R_{G}$. 

Another relevant concept in this topic is the \textit{quantum steering ellipsoid}, which is defined as the set of all Bloch vectors of the conditional states in Alice's system, see Eq.~\eqref{eq_conditional_states}, for all Bob's measurement directions in $B$ (for $y=0$)~\cite{jevtic2015}:
$$\left\{\vec{t}_{A|0}(\vb) =\frac{\tA+\C\vb}{1+\vb\cdot\tB} : \vb\in\BlochSphere\right\},$$
where $\vec{t}_{A|y}(\vb)$ is given in Eq.~\eqref{eq_conditional_bloch_vect}. The center of this 
ellipsoid is in Eq.~\eqref{eq_center_ellipsoid}:
\begin{align*}
    \centerSteeringEllipsoid = \frac{\tA-\C\tB}{1-t_B^2}.
\end{align*}

\section{Quantum key distribution protocol basics}\label{app_qkd}
Quantum key distribution protocols aim to establish a shared secret key between two parties (Alice and Bob) using quantum systems and a public classical communication channel. These protocols typically involve two stages: 1) The \textit{quantum transmission phase}, where shared bit strings are generated through quantum state preparation and measurements; and 2) the \textit{Security Analysis}, involving classical post-processing to generate a secure key from these shared strings.

Entanglement-based QKD protocols utilize pre-distributed, maximally entangled Bell states $\rho_{AB}^{\text{Bell}}$ in the quantum transmission phase. Alice and Bob generate the shared bit strings by performing local measurements on their respective subsystems. 

The Devetak-Winter rate is an important result regarding the security of arbitrary quantum key distribution protocols. It establishes a lower bound for the asymptotic secure key rate $\devetakWinterK$ generated by an arbitrary measurement $R$ on subsystem $A$~\cite{devetakwinter2005}:
\begin{align}
    \devetakWinterK\geq \Entropy(R|E)-\Entropy(R|B).
\end{align}
For the security analysis, the resource state $\rho_{AB}$, i.e. the state distributed to Alice and Bob and that is locally measured to obtain the shared bit strings, is assumed to be entirely generated by Eve (i.e. the eavesdropper), namely, the global system before the measurements occupy the --pure-- state $\rho_{ABE}$. The terms $\Entropy(R|E)$ and $\Entropy(R|B)$ represent the classical-quantum conditional von Neumann entropies, quantifying the information Eve and Bob, respectively, have about Alice's measurement outcomes. 


\section{Demonstration Result \ref{res:rate_Bound}}\label{app:Rate_bound_demonstration}
The main inequality to demonstrate Result~\ref{res:rate_Bound} is~\cite{berta2010}:
\begin{align*}
    \devetakWinterK\geq \Entropy(R|E)-\Entropy(R|B) \geq 1-\Entropy(R|R')-\Entropy(S|S'),
\end{align*}
where $R$ and $R'$, and $S$ and $S'$, are two arbitrary pairs of measurements ($R$ and $R'$ are performed over subsystem $A$, and $S$ and $S'$ over $B$), being additionally $R$ and $S$ are incompatible measurements. 

Let us take thus $R=\mathcal{O}(\va_1)$, $S=\mathcal{O}(\va_2)$, $R'=\mathcal{O}(\vb_1)$, and $S'=\mathcal{O}(\vb_2)$, as given by Eqs.~\eqref{eq_projectorsA} and \eqref{eq_projectorsB}. Given that for each measurements pair holds $\Entropy(X|X')\leq \BinaryEntropy{\BayesRisk(X,X')}$~\cite{devroye2013, cover_thomas2006}, we have that:
\begin{align}
    \label{eq:devetak_bound_measurements}
\devetakWinterK(\va_1,\vb_1,\va_2,\vb_2)\geq 1-  \BinaryEntropy{\BayesRisk(\va_1,\vb_1)}-\BinaryEntropy{\BayesRisk(\va_2,\vb_2)}=\ratebound(\va_1,\vb_1,\va_2,\vb_2).
\end{align}
Thus, in particular:
\begin{align}\label{eq:devetak_bound_measurementsV2}    \devetakWinterK(\va_1,\va_2)=\devetakWinterK[\va_1,\boboptimalMeasurement(\va_1),\va_2,\boboptimalMeasurement(\va_2)]\geq\rateboundOpt(\va_1,\va_2)=\ratebound[\va_1,\boboptimalMeasurement(\va_1),\va_2,\boboptimalMeasurement(\va_2)].
\end{align}
This proves Eq.~\eqref{eq:devetak_winter_bb84_mod}. To demonstrate Eq.~\eqref{eq:devetak_winter_bb84_mod_vs_bb84}, we use that
$$\ratebound[\va_1,\boboptimalMeasurement(\va_1),\va_2,\boboptimalMeasurement(\va_2)]\geq \ratebound(\va_1,\vb_1,\va_2,\vb_2),$$
 for all $\vb_1$ and $\vb_2$, because of $\BayesRisk(\va_i,\vb_i)\geq \BayesRisk[\va_i,\boboptimalMeasurement(\va_i)]$ for $i=1,2$, as shown in Result~\ref{res:Bayesrisk} (Sec.~\ref{sec:BayesRisk}) --note that $\BayesRisk\in[0,1/2]$. Then, in general it holds: $\ratebound(\va_1,\vb_1,\va_2,\vb_2)\geq 1-\BinaryEntropy{\QBER(\va_1,\vb_1)}-\BinaryEntropy{\QBER(\va_2,\vb_2)}$. Therefore:
 $$\rateboundBB\leq\ratebound(\vec{k},\vec{k},\vec{i},\vec{i})\leq\ratebound[\vec{k},\boboptimalMeasurement(\vec{k}),\vec{i},\boboptimalMeasurement(\vec{i})]\leq \max_{\va_1\perp\va_2}\ratebound[\va_1,\boboptimalMeasurement(\va_1),\va_2,\boboptimalMeasurement(\va_2)]\leq \devetakWinterK(\va_1^*,\va_2^*),$$
 where 
 $$\max_{\va_1\perp\va_2}\ratebound[\va_1,\boboptimalMeasurement(\va_1),\va_2,\boboptimalMeasurement(\va_2)] = \ratebound[\va_1^*,\boboptimalMeasurement(\va_1^*),\va_2^*,\boboptimalMeasurement(\va_2^*)].$$

\section{Top Quarks}\label{app:TopQuarks}

Top quark pairs ($t\bar{t}$) are ideal candidates within the Standard Model for studying the measurement of their spin correlations through the kinematic distribution of their decay products. Recently, several works have addressed this topic by first determining the full quantum state of a $t\bar{t}$ pair, $\rho(\vec{t}_{A}, \vec{t}_{B}, \textbf{C})$. Such research has found that this production density matrix describes a \textit{Bell diagonal state}, which allows for the exploration of concepts typically associated with quantum information, including quantum entanglement, quantum discord, and steering \cite{afik_quantum_2022, afik_quantum_2023}. 

\begin{figure}[t]
     \centering
         \includegraphics[width=0.35\textwidth]{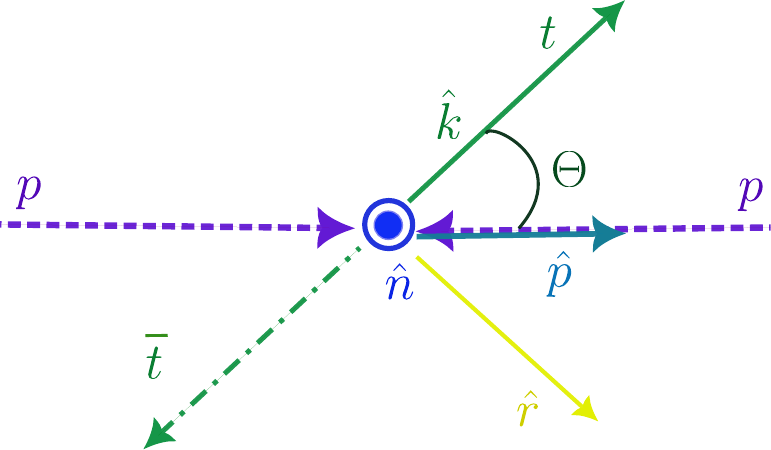}
         \caption{Orthonormal helicity basis. Here, $\hat{p}$ is a unit vector along the direction of the initial beam, $\hat{k}$ is the direction of the top and $\Theta$ is the production angle of the top quark for the beam axis ($\cos \Theta= \hat{k} \cdot \hat{p}$). The vector $\hat{n}$ is perpendicular to the $\{\hat{k}, \hat{p}\}$ plane and $\hat{r}= (\hat{p}-\cos \Theta \hat{k})/ \sin \Theta$ is the vector orthogonal to $\hat{k}$ within the $\{\hat{k}, \hat{p}\}$ plane \cite{afik_quantum_2022}.}
         \label{fig:helicitybasis}
    \end{figure}

Within the framework of quantum chromodynamics (QCD), at leading-order (LO) perturbation theory, the production of $t\bar{t}$ pairs can be the result of two initial processes ($I$): a light quark-antiquark ($I=q \bar{q}$) pair, or a gluon ($I=gg$) pair.
\begin{align}
    q + \bar{q} \rightarrow t + \bar{t}, \nonumber \\
    g + g \rightarrow t + \bar{t}.   
\end{align}  
These decays are analyzed in the \textit{helicity basis} $\{ \hat{k}, \hat{n}, \hat{r} \}$, an orthonormal basis defined in the center-of-mass (c.m.) frame of the collision (see Fig. \ref{fig:helicitybasis}). Within this basis, the production density matrix $\rho^{I}(\beta, \Theta)$ only depends on the center-of-mass energy $\beta$ (or, equivalently, on the invariant mass $M_{t\bar{t}}$) and the production angle $\cos \Theta$, parameters that govern the decay kinematics.
In addition, at LO QCD $\rho^{I}(\beta, \Theta)$ is unpolarized, meaning that the Bloch vectors are null ($\vec{t}^{I}_{A,B}=0$) and having a symmetric correlation matrix which is diagonal in the $\hat{n}$-direction. This allows it to be diagonalized by an appropriate rotation in the $\{ \hat{k}, \hat{r} \}$ plane, yielding eigenvalues $\{ C^{I}_{+}, C^{I}_{nn}, C^{I}_{-} \}$. We conclude that the production density matrix of a $t\bar{t}$ pair is a \textit{Bell diagonal state}.  

The coefficients of the diagonal correlation matrix $\{ C^{I}_{+}(\beta, \Theta), C^{I}_{nn}(\beta, \Theta), C^{I}_{-}(\beta, \Theta) \}$ in the helicity basis are given by
\begin{equation}\label{eq_corr_elements}
    C^{I}_{\pm}=\frac{C^{I}_{kk}+C^{I}_{rr}}{2} \pm \sqrt{ \left( \frac{C^{I}_{kk}-C^{I}_{rr}}{2} \right)^{2}+ C_{kr}^{I^{2}}},
\end{equation}
with $C_{ij}=\tilde{C}_{ij}/ \tilde{A}$ defined for each process as follows:
\begin{itemize}
     \item $q\bar{q}$ processes  
     \begin{align}
    \tilde{A}^{q\bar{q}}&=F_{q}(2-\beta^{2} \sin^{2} \Theta), \qquad   \nonumber  \\    
    \tilde{C}^{q\tilde{q}}_{rr}&= F_{q}(2-\beta^{2})\sin^{2} \Theta,  \nonumber \\
    \tilde{C}^{q\tilde{q}}_{nn}&= -F_{q}\beta^{2}\sin^{2} \Theta,  \nonumber \\
    \tilde{C}^{q\tilde{q}}_{kk}&= F_{q}\left[2-(2-\beta^{2})\sin^{2}\Theta \right],  \nonumber \\
    \tilde{C}^{q\tilde{q}}_{rk}&=\tilde{C}^{q\tilde{q}}_{kr}= F_{q}\sqrt{1-\beta^{2}} \sin 2\Theta, \nonumber \\
    F_{q}&= \frac{1}{18}.\label{eq_qq_process}
\end{align}
    \item $gg$ processes 
    \begin{align}
    \tilde{A}^{gg}&=F_{g}\left[1+2\beta^{2}\sin^{2}\Theta-\beta^{4}(1+\sin^{4} \Theta)     \right], \nonumber \\  
    \tilde{C}^{gg}_{rr}&=- F_{g}\left[1-\beta^{2}(2-\beta^{2})(1+\sin^{4} \Theta)   \right],  \nonumber \\
    \tilde{C}^{gg}_{nn}&= -F_{g} \left[1-2\beta^{2}+\beta^{4}(1+\sin^{4} \Theta)  \right],  \nonumber \\
    \tilde{C}^{gg}_{kk}&= -F_{g}\left[1-\beta^{2} \frac{\sin^{2}2\Theta}{2}-\beta^{4}(1+\sin^{4} \theta) \right],  \nonumber \\
    \tilde{C}^{gg}_{rk}&=\tilde{C}^{gg}_{kr}= F_{g}\sqrt{1-\beta^{2}} \beta^{2} \sin 2\Theta \sin^{2}\Theta,  \nonumber \\
    F_{g}&= \frac{7+9\beta^{2} \cos^{2} \Theta}{192(1-\beta^{2} \cos^{2} \Theta)^{2}}.\label{eq_gg_process}
\end{align}
 \end{itemize}
However, in a realistic scenario, the production of $t\bar{t}$ pairs occurs predominantly through two main types of realistic processes: $pp$ and $p\bar{p}$ collisions. To properly analyze these interactions, it is crucial to acknowledge the composite nature of protons and antiprotons. Composed of quarks and gluons (partons), they are classified as \textit{hadronic particles} (baryons). Thus, the production of $t\bar{t}$ pairs via $pp$ or $p\bar{p}$ collisions at certain energy scale depends on the contribution of each parton, which is described through its parton distribution function (PDF).

Therefore, for a given top quark direction $\hat{k}$ and center-of-mass energy $\sqrt{s}$, the correlations $\C(M_{t\bar{t}}, \Theta, \sqrt{s})$ of $\rho(M_{t\bar{t}}, \Theta, \sqrt{s})$ are derived as a mixture of each initial state $I=q\bar{q},gg$ of the total hadronic process as follows:
\begin{equation}
    C_{ij}(M_{t\bar{t}}, \Theta, \sqrt{s})=\sum_{I=q\bar{q},gg} w_{I}(M_{t\bar{t}}, \Theta, \sqrt{s})C_{ij}^{I}(M_{t\bar{t}}, \Theta),
\end{equation}
where the weights $w_{I}(M_{t\bar{t}}, \Theta, \sqrt{s})$, representing the probability of producing each partonic quantum state $\rho^{I}(M_{t\bar{t}}, \hat{k})$, are directly computed from the respective parton luminosity functions $L_{I}(M_{t\bar{t}}, \sqrt{s})$~\cite{ball2015}, as shown below:
\begin{equation}
    w_{I}(M_{t\bar{t}}, \Theta, \sqrt{s})=\frac{L_{I}(M_{t\bar{t}}, \sqrt{s})\tilde{A}^{I}(M_{t\bar{t}}, \Theta )}{\sum_{J}L_{J}(M_{t\bar{t}}, \sqrt{s})\tilde{A}^{J}(M_{t\bar{t}}, \Theta)},
\end{equation}
such that $w_{q\bar{q}}+w_{gg}=1$. 

\end{document}